\documentclass[twocolumn,aps,showpacs,prb,tightenlines,amsmath,amssymb,superscriptaddress, citeautoscript]{revtex4-1}

\usepackage{graphicx}

\usepackage{dcolumn}

\usepackage{array}
\newcolumntype {L}{>{$}l<{$}}             
\newcolumntype {C}{>{$}c<{$}}             
\newcolumntype {R}{>{$}r<{$}}             
\newcolumntype {s}[1]{@{\hspace*{#1}}}    

\newcommand* {\ket}[1]{| {#1} \rangle}
\newcommand* {\braket}[1]{\langle {#1} \rangle}

\usepackage{bm}

\newcommand* {\vek}[1]{{\bm{#1}}}
\newcommand* {\vekc}[1]{{\bm{\mathcal{#1}}}}
\newcommand* {\kk}{\vek{k}}
\newcommand* {\tvek}[2][c]{\left( \begin{array}{s{0.15em}#1s{0.15em}}
     #2\end{array} \right)}

\usepackage{color}
\definecolor{green}{rgb}{0.0,0.60,0.0}

\usepackage[normalem]{ulem}

\usepackage[english]{babel}

\begin{document}

\title{Asymmetric $\vek{g}$ tensor in low-symmetry two-dimensional hole systems}
\author{C.\ Gradl}
\thanks{These authors contributed equally}
\affiliation{Institut f\"ur Experimentelle und Angewandte Physik,
Universit\"at Regensburg, D-93040 Regensburg, Germany}
\author{R.\ Winkler}
\thanks{These authors contributed equally}
\affiliation{Department of Physics, Northern Illinois University, DeKalb, Illinois 60115, USA}
\author{M.\ Kempf}
\affiliation{Institut f\"ur Experimentelle und Angewandte Physik,
Universit\"at Regensburg, D-93040 Regensburg, Germany}
\author{J.\ Holler}
\affiliation{Institut f\"ur Experimentelle und Angewandte Physik,
Universit\"at Regensburg, D-93040 Regensburg, Germany}
\author{D.\ Schuh}
\affiliation{Institut f\"ur Experimentelle und Angewandte Physik,
Universit\"at Regensburg, D-93040 Regensburg, Germany}
\author{D.\ Bougeard}
\affiliation{Institut f\"ur Experimentelle und Angewandte Physik,
Universit\"at Regensburg, D-93040 Regensburg, Germany}
\author{A.\ Hern\'andez-M\'inguez}
\affiliation{Paul-Drude-Institut f\"ur Festk\"orperelektronik, Leibniz-Institut im Forschungsverbund Berlin e.V., D-10117 Berlin, Germany}
\author{K.\ Biermann}
\affiliation{Paul-Drude-Institut f\"ur Festk\"orperelektronik, Leibniz-Institut im Forschungsverbund Berlin e.V., D-10117 Berlin, Germany}
\author{P. V.\ Santos}
\affiliation{Paul-Drude-Institut f\"ur Festk\"orperelektronik, Leibniz-Institut im Forschungsverbund Berlin e.V., D-10117 Berlin, Germany}
\author{C.\ Sch\"uller}
\affiliation{Institut f\"ur Experimentelle und Angewandte Physik,
Universit\"at Regensburg, D-93040 Regensburg, Germany}
\author{T.\ Korn}
\email{tobias.korn@ur.de}
\affiliation{Institut f\"ur Experimentelle und Angewandte Physik,
Universit\"at Regensburg, D-93040 Regensburg, Germany}

\date{\today}

\begin{abstract}
The complex structure of the valence band in many semiconductors leads to multifaceted and unusual properties for spin-3/2 hole systems compared to common spin-1/2 electron systems. In particular, two-dimensional hole systems show a highly anisotropic Zeeman interaction. We have investigated this anisotropy in GaAs/AlAs quantum well structures both experimentally and theoretically. By performing time-resolved Kerr rotation measurements, we found a non-diagonal tensor $\vek{g}$ that manifests itself in unusual precessional motion as well as distinct dependencies of hole spin dynamics on the direction of the magnetic field $\vek{B}$. We quantify the individual components of the tensor $\vek{g}$ for [113]-, [111]- and [110]-grown samples.  We complement the experiments by a comprehensive theoretical study of Zeeman coupling in in-plane and out-of-plane fields $\vek{B}$.  To this end, we develop a detailed multiband theory for the tensor $\vek{g}$.  Using perturbation theory, we derive transparent analytical expressions for the components of the tensor $\vek{g}$ that we complement with accurate numerical calculations based on our theoretical framework. We obtain very good agreement between experiment and theory.  Our study demonstrates that the tensor $\vek{g}$ is neither symmetric nor antisymmetric.  Opposite off-diagonal components can differ in size by up to an order of magnitude.  The tensor $\vek{g}$ encodes not only the Zeeman energy splitting but also the direction of the axis about which the spins precess in the external field $\vek{B}$.  In general, this axis is not aligned with $\vek{B}$.  Hence our study extends the general concept of optical orientation to the regime of nontrivial Zeeman coupling.
\end{abstract}


\maketitle

\section{Introduction} \label{sec:introduction}

Spin-dependent phenomena in semiconductor heterostructures have been studied intensively in recent years with the prospect of enabling spintronics and quantum information applications \cite{awschalom2002, winkler2003, zutic2004, fabian2007, dyakonov2008, wu2010, trifunovic2012}.  Many studies have investigated the dynamics of electron systems in direct-gap semiconductors like GaAs. Here, bulk systems as well as nanostructures ranging from quantum wells (QWs) to wires and dots have been covered. While electrons in the conduction band of semiconductors such as GaAs are $s$-like with an effective spin 1/2, the $p$-like character of holes in the valence band gives rise to an effective spin 3/2 that offers more complex and novel spin-dependent characteristics. To be able to observe these special properties in detail, sufficiently long hole-spin coherence times are required, which are not accessible in bulk systems \cite{hilton02, PhysRevB.82.115205}. In recent years, these coherence times have been pushed in QWs from picoseconds to almost 100~ns using special sample designs and low temperatures \cite{damen1991, marie1999, syperek2007, kugler2009, korn2010, kugler2011}. In quantum dot systems, equally impressive progress has been achieved in studying long-lived hole spin dynamics \cite{Muto_hole_g_Dot_APL07, heiss07, Seidl2008, Crooker_HoleSpinNoise_PRL10, Schwan11, Oestreich_singleHole_PRL14, PhysRevB.93.035311}. Hole spins are attractive candidates for quantum information processing schemes, as the contact hyperfine interaction with nuclei is suppressed due to the $p$-like character of the hole wave functions \cite{fischer:155329, PhysRevB.80.235320, PhysRevB.86.085319}.  One of the unique properties of low-dimensional hole systems is a highly anisotropic Zeeman interaction characterized by a tensor $\vek{g}$ instead of a scalar effective $g$ factor.

The Zeeman interaction in low-dimensional electron systems is rather independent of the particular orientation of the magnetic field $\vek{B}$ \cite{fang1968}. In two-dimensional (2D) systems, the effective electron $g$ factor predominantly depends on the width of the QW, i.e., the penetration of the electron wave function into the barriers and the quantization energy in the QW \cite{snelling1991, hannak1995, yugova2007, shchepetilnikov2013}. The anisotropy between the in-plane and out-of-plane direction is relatively small except for narrow QWs \cite{ivchenko1992, kalevich1992, pfeffer2006, shchepetilnikov2013, sandoval2016}. A reduction of the symmetry of the system leads to an anisotropic Zeeman splitting of electrons for different in-plane directions of $\vek{B}$ \cite{huebner2011, kalevich1993, eldridge2011, nefyodov2011, english2013}. On the one hand, low-symmetry growth directions cause a slight anisotropy of the diagonal components $g_{xx}\neq g_{yy}$ \cite{huebner2011}. On the other hand, due to asymmetric band-edge profiles small off-diagonal components $g_{xy}=g_{yx}\neq0$ arise while $g_{xx}=g_{yy}$ \cite{kalevich1993, eldridge2011, nefyodov2011, english2013, footnote1}.

Remarkably, the $g$ factor in spin-3/2 hole systems shows a strong
dependence on the direction of $\vek{B}$. The Zeeman splitting can
differ by an order of magnitude for the in-plane and out-of-plane
directions of $\vek{B}$ \cite{kesteren1990, wimbauer1994, korn2010,
arora2013, simion2014, terentev2017, miserev2017}. (See also
Refs.~\onlinecite{danneau06, csontos08, koduvayur08, chen10,
komijani13, srinivasan16} for related work on quantum wires.)
Furthermore, it depends sensitively on the growth direction of the
QW \cite{winkler2000, winkler2003}. For low-symmetry growth
directions the Zeeman splitting becomes highly anisotropic for
different directions of $\vek{B}$ in the QW plane \cite{winkler2000,
gradl2014}. Off-diagonal elements $g_{zx}$ of the hole $\vek{g}$
tensor result in peculiar properties such as a non-collinear
paramagnetism, where an in-plane magnetic field gives rise to an
out-of-plane spin polarization \cite{winkler2003, winkler2008}.
However, so far these theoretical predictions of off-diagonal
$\vek{g}$-tensor components have only been experimentally verified
qualitatively \cite{yeoh2014}.

Traditional experimental techniques to study the Zeeman interaction including electron paramagnetic resonance and magneto-photoluminescence can only detect the Zeeman energy splitting which is governed by the symmetric tensor $\vek{G} \equiv \vek{g}^\dagger \cdot \vek{g}$ [see Eq. (\ref{eq:Gdefinition}) below] \cite{abragam1970, grachev1987, mcgavin1990, snelling1992, findeis2000}. The tensor $\vek{g}$, on the other hand, determines also the direction of the spin $\vek{S}$, which generally is not aligned with the external field $\vek{B}$.  In our fully quantitative study, we determine both experimentally and theoretically the complete tensor $\vek{g}$ for several low-symmetry 2D GaAs hole systems with very good agreement between experiment and theory.  Our work demonstrates that the tensor $\vek{g}$ is, in general, neither symmetric ($\vek{g} \ne \vek{g}^\dagger$) nor antisymmetric ($\vek{g} \ne - \vek{g}^\dagger$). Opposite off-diagonal components can differ in size by up to an order of magnitude.

In our experiments, we use time-resolved Kerr rotation (TRKR) to extract the Larmor spin precession frequency for a large range of in-plane and out-of-plane magnetic fields that allows us to extract the diagonal and off-diagonal components of the tensor $\vek{G}$ characterizing the Zeeman energy splitting. Our sample design enables us to study low-symmetry QW orientations, while providing sufficient hole spin coherence times to accurately determine the precession frequencies. The precise control of the QW symmetry, which would be hard to realize in, e.g., self-organized quantum dot systems, allows us to map out the parameter space for the tensor $\vek{G}$ as a function of crystal orientation.  As the Kerr effect (for our measurement geometry) is only sensitive to an out-of-plane spin polarization, this effect allows us to probe also the precessing spin vector $\vek{S}$ itself (as a function of the direction of $\vek{B}$).  Here the signature is a Kerr signal that contains an oscillatory and a non-oscillatory component.  In this way, we can also determine the components of the tensor $\vek{g}$ itself, thus extending the general concept of \emph{optical orientation} \cite{meier84} (exploiting the fact that electron spin and orbital angular momenta can be probed and manipulated with optical techniques) to the regime of nontrivial Zeeman coupling.  In this regime, the light field and the external magnetic field provide two \emph{independent} ways to address the vector of spin angular momentum.

We complement the experiments by a comprehensive theoretical study of the Zeeman interaction in in-plane and out-of-plane fields $\vek{B}$. To this end, we begin with a thorough symmetry analysis for the tensor $\vek{g}$ based on the theory of invariants which provides detailed predictions on nonzero and vanishing components of $\vek{g}$ as a function of the crystallographic growth direction, thus illustrating the concept of \emph{invariant tensors}.  Then we develop a detailed multiband theory for the tensor $\vek{g}$. Using perturbation theory, we derive transparent analytical expressions for the components of the tensor $\vek{g}$ that we complement with accurate numerical calculations based on our theoretical framework. We obtain very good agreement between experiment and theory.

Besides its fundamental importance, a detailed knowledge of the tensor $\vek{g}$ and its potential asymmetry is especially important for spintronics and quantum information applications where the manipulation of the spin vector $\vek{S}$ plays a central role \cite{loss1998, awschalom2002}. In addition to the apparent magnetic control of the spin via the $\vek{g}$ tensor, it has also been shown that the $\vek{g}$ tensor is crucial for a possible electric control of single spins in qubits \cite{andlauer2009, roloff2010, PhysRevB.84.195403}.

\section{Theoretical Considerations}
\label{sec:Theo}

We are interested in the Zeeman interaction
\begin{equation}
\label{eq:zeeman}
H_Z = \frac{\mu_\mathrm{B}}{2}
\, \vek{\sigma} \cdot \vek{g} \cdot \vek{B}
\end{equation}
in quasi-2D systems.  Here $\mu_\mathrm{B}$ is the Bohr magneton,
and $\vek{g}$ is generally a second-rank tensor (a $3\times 3$
matrix) that couples the spin operator
$\vek{S} = (\hbar/2) \vek{\sigma}$ to the external magnetic
field~$\vek{B}$.  The quantity $\vek{\sigma}$ denotes the vector of
Pauli matrices.

\subsection{Symmetry of the tensor $\vek{g}$}
\label{sec:sym}

The theory of invariants \cite{bir1974, winkler2003} allows one to
determine the structure of the tensor $\vek{g}$ using only symmetry
arguments.  This analysis is summarized in Table~\ref{tab:invar}.
For symmetric GaAs (zincblende) quasi-2D systems with growth
direction $[mmn]$ ($m$, $n$ integer), the point group is in general
$C_s$, while the high-symmetry directions [001], [111], and [110]
yield $D_{2d}$, $D_{3d}$, and $C_{2v}$, respectively.  The
irreducible representations (IRs) $\Gamma_i$ of the HH states for
each of the respective double groups are also indicated in this
table.  These IRs are labeled following Koster \emph{et al.}
\cite{koster1963} For HH systems with growth direction [111] (point
group $D_{3d}$), time-reversal symmetry requires one to combine the
one-dimensional complex conjugate IRs $\Gamma_5$ and $\Gamma_6$; and for growth
direction $[mmn]$ (point group $C_s$), one has to combine $\Gamma_3$
and $\Gamma_4$.

\begin{table}
  \caption{Irreducible tensor components $k_i$, $B_i$ and
  corresponding basis matrices $\sigma_i$ for HH systems with point
  group symmetries $D_{2d}$, $D_{3d}$, $C_{2v}$, and $C_{s}$,
  corresponding to quasi-2D systems with growth directions [001],
  [111], [110], and $[mmn]$, respectively.  For these symmetries,
  the HH systems transform according to the irreducible
  representations $\Gamma_i$ indicated in brackets.  The invariants that can be
  formed by combining tensor components with basis matrices and the
  resulting invariant tensors $\vek{g}$ and $\vek{\gamma}$ are also
  listed. Irreducible representations $\Gamma_i$ for the point
  groups are labeled following Koster \emph{et~al.} \cite{koster1963}
  \label{tab:invar}}
  \centering
\renewcommand{\arraystretch}{1.2}
\newcommand{\invtensor}{{\renewcommand{\arraystretch}{0.90}\begin{tabular}{l}
  invariant\\ tensors\end{tabular}}}
\begin{tabular*}{\columnwidth}{Cs{2em}L} \hline\hline
  \multicolumn{2}{C}{D_{2d} \; (\mbox{HH:}~\Gamma_7) \quad [001]} \\ \hline
  \Gamma_2 & B_z; \sigma_z \\
  \Gamma_4 & k_z \\
  \Gamma_5 & k_x, k_y; B_y, B_x; \sigma_y, -\sigma_x \\
  \mbox{invariants} & \sigma_x B_x - \sigma_y B_y; \sigma_z B_z;
  \sigma_y k_x - \sigma_x k_y \\
  \invtensor
  & \tvek[ccc]{g_{xx} & 0 & 0 \\ 0 & - g_{xx} & 0 \\ 0 & 0 & g_{zz}},
  \tvek[ccc]{0 & - \gamma_{yx} & 0 \\ \gamma_{yx} & 0 & 0 \\ 0 & 0 & 0}
  \\ \hline
  \multicolumn{2}{C}{D_{3d} \; (\mbox{HH:}~\Gamma_5 \oplus \Gamma_6)
  \quad [111]} \\ \hline
  \Gamma_1 & k_z; \sigma_y \\
  \Gamma_2 & B_z; \sigma_x; \sigma_z \\
  \Gamma_3 & k_x, k_y; B_y, - B_x \\
  \mbox{invariants} & \sigma_x B_z; \sigma_z B_z; (\sigma_y k_z) \\
  \invtensor
  & \tvek[ccc]{0 & 0 & g_{xz} \\ 0 & 0 & 0 \\ 0 & 0 & g_{zz}},
  \tvek[ccc]{0 & 0 & 0 \\ 0 & 0 & (\gamma_{yz}) \\ 0 & 0 & 0}
  \\ \hline
  \multicolumn{2}{C}{C_{2v} \; (\mbox{HH:}~\Gamma_5) \quad [110]} \\ \hline
  \Gamma_1 & k_x \\
  \Gamma_2 & k_y; B_z; \sigma_z \\
  \Gamma_3 & B_x; \sigma_x \\
  \Gamma_4 & k_z; B_y; \sigma_y \\
  \mbox{invariants} & \sigma_x B_x; \sigma_y B_y; \sigma_z B_z;
  (\sigma_y k_z); \sigma_z k_y \\
  \invtensor
  & \tvek[ccc]{g_{xx} & 0 & 0 \\ 0 & g_{yy} & 0 \\ 0 & 0 & g_{zz}},
  \tvek[ccc]{0 & 0 & 0 \\ 0 & 0 & (\gamma_{yz}) \\ 0 & \gamma_{zy} & 0}
  \\ \hline
  \multicolumn{2}{C}{C_s \; (\mbox{HH:}~\Gamma_3 \oplus \Gamma_4)
  \quad [mmn]} \\ \hline
  \Gamma_1 & k_x; k_z; B_y; \sigma_y \\
  \Gamma_2 & k_y; B_x; B_z; \sigma_x; \sigma_z \\
  \mbox{invariants} & \sigma_x B_x; \sigma_x B_z; \sigma_y B_y;
  \sigma_z B_x; \sigma_z B_z; \\
  & \sigma_x k_y; \sigma_y k_x; (\sigma_y k_z); \sigma_z k_y \\
  \invtensor
  & \tvek[ccc]{g_{xx} & 0 & g_{xz} \\ 0 & g_{yy} & 0 \\ g_{zx} & 0 & g_{zz}},
  \tvek[ccc]{0 & \gamma_{xy} & 0 \\ \gamma_{yx} & 0 & (\gamma_{yz}) \\ 0 & \gamma_{zy} & 0}
  \\ \hline\hline
\end{tabular*}
\end{table}

\begin{figure}[t]
	\begin{center}\includegraphics[width=8.5cm]{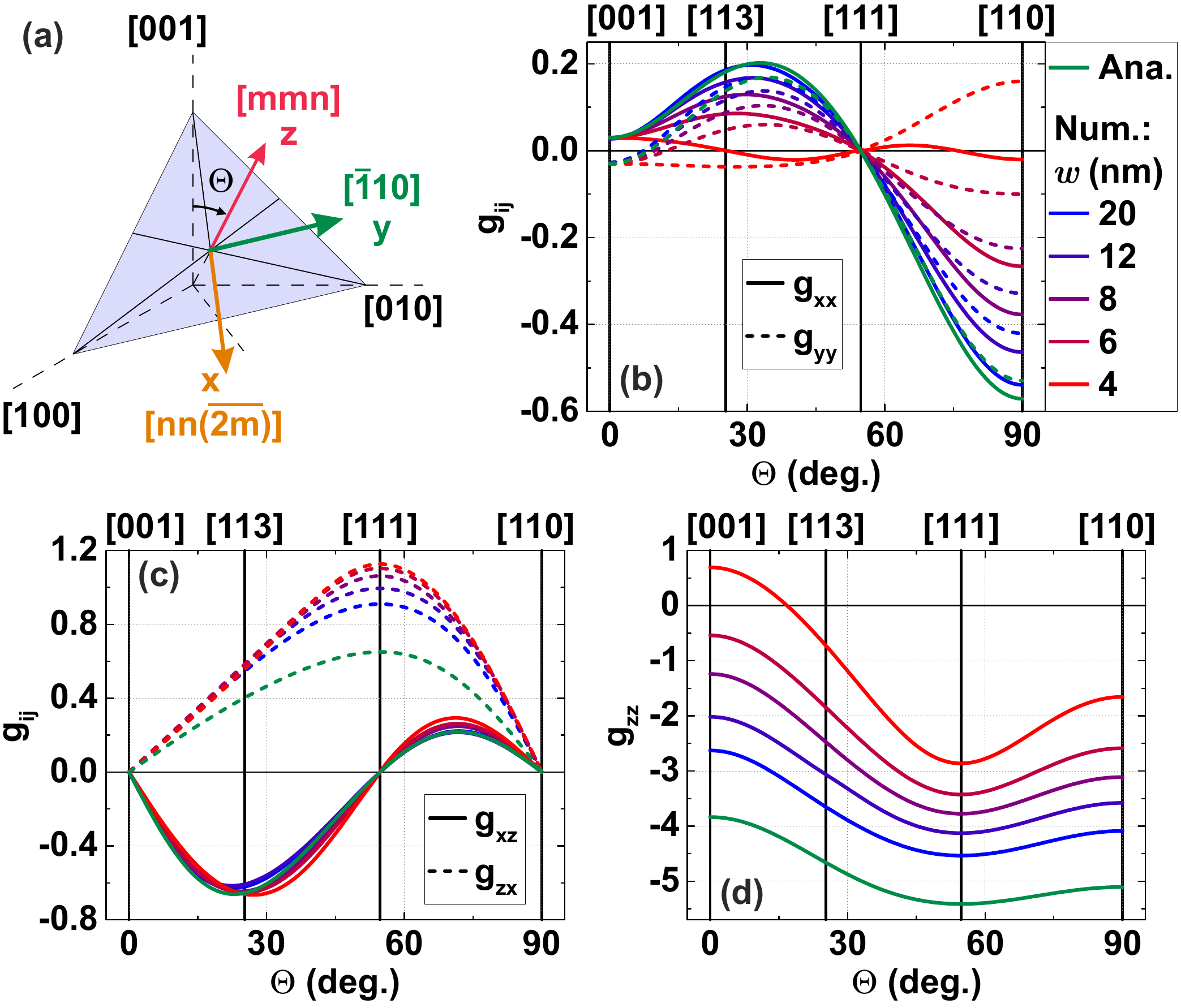}\end{center}
	\caption{(a) Coordinate system defining the angle $\theta$ as well as the crystallographic orientation of the axes $x$, $y$, and $z$. (b)-(d) Components of the tensor $\vek{g}$ for the HH1 subband in a symmetric GaAs-AlAs QW according to the analytical model and numerical calculations for different QW widths $w$. \label{fig:wellwidth}}
\end{figure}

The irreducible tensor components of the magnetic field
$\vek{B} = (B_x, B_y, B_z)$ and the corresponding basis matrices
$\sigma_i$ for HH systems are likewise listed in
Table~\ref{tab:invar}, assuming the coordinate system in
Fig.~\ref{fig:wellwidth}(a).  Here all IRs of the basis matrices
$\sigma_i$ are real so that invariants are formed by combining
tensor components with basis matrices that transform according to
the same IR.  These invariants are listed in Table~\ref{tab:invar},
too.  For reasons discussed below, we also include invariants that
can be formed from irreducible tensor components of the wave vector
$\vek{k} = (k_x, k_y, k_z)$.  Invariants proportional to the wave
vector $k_z$ are listed in brackets because these must vanish in the
quasi-2D systems discussed here.  However, all results in
Table~\ref{tab:invar} apply likewise to bulk systems where the given
point groups can be realized, e.g., by applying uniaxial strain
\cite{trebin1979}. Results for the tensor $\vek{g}$ are also valid
for suitably designed quantum wire and dot systems with the listed
point group symmetries.

The prefactor for each symmetry-allowed invariant $\sigma_i B_j$ is
$(\mu_\mathrm{B}/2) g_{ij}$ with $g_{ij} \ne 0$.  For $D_{2d}$ we
get the invariant $\sigma_x B_x - \sigma_y B_y$ (thus combining
two terms $\sigma_i B_j$) implying $g_{yy} = - g_{xx} \ne 0$.  On
the other hand, we have $g_{ij} = 0$ if $\sigma_i$ and $B_j$
transform according to different IRs so that the term $\sigma_i B_j$
is not allowed by symmetry.  Thus it follows from
Table~\ref{tab:invar} that, for the geometries considered here,
symmetry always requires $g_{xy} = g_{yx} = g_{zy} = g_{yz} = 0$.
For growth direction [111], a perpendicular magnetic field $B_z$
couples not only to the spin component $S_z$ but also to $S_x$ so
that $g_{xz} \ne 0$.  However, there is no $B$-linear Zeeman
splitting for an in-plane magnetic field implying, in particular,
$g_{zx} = 0$.  Similarly, for the general case $[mmn]$ we have
$g_{xz} \ne \pm g_{zx}$.  In the theory of invariants, this is
indicated by the fact that $\sigma_x B_z$ and $\sigma_z B_x$ are
independent, unrelated invariants, where each has its own prefactor.
Hence, the tensor $\vek{g}$ is not required by symmetry to be
symmetric or antisymmetric.  All results for the tensor $\vek{g}$ in
Table~\ref{tab:invar} are consistent with the explicit calculations
presented below.  Note also that the point group of an $[mmn]$-oriented QW remains $C_{s}$ even in the presence of both bulk inversion symmetry (giving rise to Dresselhaus~\cite{dresselhaus1955, malcher86} spin-orbit coupling) and structure inversion symmetry (giving rise to Rashba~\cite{bychkov1984, bihlmayer15} spin-orbit coupling).  Also, for a structure with point group symmetry $C_s$, electron and LH systems transform likewise according to $\Gamma_3 \oplus \Gamma_4$ so that the tensor $\vek{g}$ (and the tensor $\vek{\gamma}$ discussed below) given for $C_{s}$ in Table~\ref{tab:invar} applies to all these cases \cite{foot:groups}, consistent with earlier work \cite{huebner2011, kalevich1993, eldridge2011, nefyodov2011, english2013, footnote1}.

We want to compare these findings for the tensor $\vek{g}$ with the
well-known symmetry properties of second-rank \emph{material}
tensors connecting two \emph{field} tensors $\vek{U}$ and $\vek{V}$.
Material tensors are found to be symmetric for a number of reasons
\cite{nye1985}.  Tensors such as the permeability and permittivity are
symmetric due to the laws of equilibrium thermodynamics according to
which the electric field and the electric displacement as well as
the magnetic field and the magnetic induction are conjugate state
variables, and the differential of the free energy is exact.
Another example along these lines is the Pauli spin susceptibility
$\vek{\chi}$ of an electron gas with $g$ tensor $\vek{g}$, where
$\vek{\chi} \propto \vek{g}^\dagger \cdot \vek{g}$ so that
$\vek{\chi}$ is a symmetric tensor even if $\vek{g}$ is not
symmetric.  On the other hand, material tensors such as the electric
and thermal conductivity are symmetric due to the laws of
nonequilibrium thermodynamics (Onsager's principle).  A third reason
applies to, e.g., the tensor of thermal expansion that ``inherits''
the property of being symmetric from the strain tensor.

The tensor $\vek{g}$ appearing in Eq.\ (\ref{eq:zeeman}) does not
match any of these scenarios.  While the magnetic field $\vek{B}$
represents a thermodynamic state variable, the spin \emph{operator}
$\vek{S}$ is not its conjugate partner, so that $\vek{g}$ need not
be symmetric according to such arguments.

Interestingly, we can replace in Eq.\ (\ref{eq:zeeman}) the axial
vector $\vek{B}$ by the polar vector $\vek{k}$ (the wave vector),
giving in the Hamiltonian a term
$H_D = \vek{\sigma} \cdot \vek{\gamma} \cdot \vek{k}$ with a
second-rank tensor $\vek{\gamma}$.  Such a term $H_D$ becomes
allowed by symmetry whenever inversion symmetry is not a good
symmetry operation of the system.  In the present case of GaAs QWs,
$H_D$ characterizes the well-known $k$-linear Dresselhaus
\cite{malcher86} or Rashba \cite{bychkov1984, bihlmayer15} spin
splitting in the system.  Of course, for the quasi-2D systems
studied here the wave vector has only two Cartesian components
$(k_x, k_y)$, whereas the spin operator $\vek{S} = (S_x, S_y, S_z)$
remains a three-dimensional vector.  For example, the Dresselhaus
term $\sigma_z \gamma_{zy} k_y$ is well-known from spin relaxation
in $[110]$-grown QWs \cite{dyakonov1986, ohno1999}; yet it has no
counterpart $\sigma_y \gamma_{yz} k_z$, indicating that, generally,
the Dresselhaus tensor $\vek{\gamma}$ is likewise neither symmetric
nor antisymmetric.  Similar to the components $g_{ij}$, the nonzero
elements $\gamma_{ij}$ can be identified using the theory of
invariants.  For the geometries considered here, the resulting
second-rank tensors $\vek{\gamma}$ are also listed in
Table~\ref{tab:invar}.  For both $\vek{g}$ and $\vek{\gamma}$ each
nonzero element of these tensors represents the prefactor for an
independent invariant in the Hamiltonian (unless indicated, e.g.,
for $[001]$-grown QWs, where the invariant expansion implies
$g_{yy} = - g_{xx}$ and $\gamma_{yx} = - \gamma_{xy}$).  The physics
contained in the second-rank tensors $\vek{g}$ and $\vek{\gamma}$ is
thus very different from material tensors \cite{nye1985} such as
permeability, permittivity, and electric conductivity.  Hence we
call $\vek{g}$ and $\vek{\gamma}$ \emph{invariant tensors}.  These
tensors are, in general, neither symmetric nor antisymmetric.  Also,
while material tensors $\vek{M}$ generally have $\det \vek{M} \ne 0$
(Ref.~\onlinecite{nye1985}), we see in Table~\ref{tab:invar} that
invariant tensors $\vek{I}$ may be required by symmetry to have
$\det \vek{I} = 0$.  The latter property is also known for the
symmetric second-rank gyration tensor. \cite{nye1985}

\subsection{Multiband theory for the tensor $\vek{g}$}

In the following we want to derive a general and quantitative
multiband theory for the tensor $\vek{g}$ in quasi-2D systems.  The
motion of Bloch electrons in a magnetic field is characterized by
the kinetic wave vector
$\kk = \bm{\mathfrak{k}} + (e/\hbar) \, \vek{A}$, where
$\bm{\mathfrak{k}} = -i\nabla$ represents the canonical wave vector
and $\vek{A}$ is the vector potential for the magnetic field
$\vek{B} = \nabla \times \vek{A}$.  The components of $\vek{k}$ are
characterized by the commutator relation
\begin{equation}
\label{eq:k_com}
\kk \times \kk = \frac{e}{i\hbar} \vek{B},
\end{equation}
independent of the choice for the vector potential $\vek{A}$
corresponding to $\vek{B}$.  The general Zeeman interaction
(\ref{eq:zeeman}) is suggestive of a spin-1/2 electron system.  Yet
we emphasize that it is equally valid for, e.g., 2D HH systems that
derive from bulk hole systems characterized by an effective spin
3/2.  In all these cases we obtain for $B=0$ a two-fold degeneracy
that can be parameterized by the Pauli matrices $\sigma_i$.  Here
the eigenstates $\ket{\pm}$ of $\sigma_z$ can be defined by the
condition that they are excited by $\pm$ polarized light,
\emph{independent} of the Zeeman interaction.  It is this aspect
which allows us to determine all components of the tensor $\vek{g}$
from our experiments, as discussed in more detail below.

We assume that the 2D system is described by the $N\times N$
multiband Hamiltonian $H$.  The operator $H$ may represent, e.g.,
the $4 \times 4$ Luttinger Hamiltonian or the $8 \times 8$ Kane
Hamiltonian as discussed below.  We decompose
\begin{equation}
\label{eq:H_decomp}
H = H_0 + \vek{H}'_B \cdot \vek{B} + H'_k
\equiv H_0 + H' .
\end{equation}
Here $H_0$ defines the (exact) subband states $\ket{\alpha}$ at
$\mathfrak{k}_\| = B = 0$, i.e.,
$H_0 \ket{\alpha} = E_\alpha \ket{\alpha}$.  The remainder defines
the perturbation $H'$.  The term $\vek{H}'_B \cdot \vek{B}$ denotes
the Zeeman terms due to $\kk \cdot \vek{p}$ couplings to remote
bands outside the $N$-dimensional space.  Finally, $H'_k$ denotes
the terms proportional to the in-plane wave vector
$\kk_\| \equiv (k_x,k_y)$.  By definition, it contains only
symmetrized products of $\kk_\|$ and $k_z$.  In general, we get
\begin{equation}
\label{eq:g_tensor_parts}
\vek{g} = \vek{g}_0 + \Delta\vek{g} ,
\end{equation}
where $\vek{g}_0$ stems from first-order perturbation theory for
$\vek{H}'_B \cdot \vek{B}$ and $\Delta\vek{g}$ is due to first- or
second-order perturbation theory for $H'_k$.

In the limit $\kk_\|, \vek{B} \rightarrow 0$, the subband states
$\ket{\alpha}$ are twofold degenerate, i.e., the Zeeman term
(\ref{eq:zeeman}) acts in the subspace defined by the
spin-degenerate pairs of states $\{\ket{\alpha}, \ket{\alpha'}\}$,
taken to be eigenstates of $\sigma_z$ as discussed above.
We may decompose
\begin{equation}
H_0 \equiv \tilde{H}_0 + H_0' ,
\end{equation}
Here $\tilde{H}_0$ is the band-diagonal part of $H_0$.  The
off-diagonal part $H_0'$ gives rise to a nontrivial spinor structure
of the eigenstates $\{ \ket{\alpha}, \ket{\alpha'} \}$ of $H_0$.  As
discussed in more detail below, only when the terms $H_0'$ are
included in $H_0$, lowest-order (i.e., first- or second-order)
perturbation theory is sufficient for exact results on Zeeman
splitting.

Often $H_0'$ is small compared with $\tilde{H}_0$ so that we can
include $H_0'$ in the perturbation $H'$.  The unperturbed
eigenstates of $\tilde{H}_0$ are band-diagonal [e.g., either heavy
hole (HH) or light hole (LH)].  We will see below that in this case
$H_0'$ gives rise to additional terms in the perturbative expansion
(including mixed terms depending also on other parts of $H'$) that
are otherwise hidden in the definition of the proper eigenstates of
$H_0$.  Often it is illuminating to make the interplay of these
terms explicit.  Yet this also implies that starting from
$\tilde{H}_0$ we generally need one extra order of perturbation
theory for $\vek{g}$ to construct nontrivial wave functions that
incorporate band mixing.  This approach was used in
Ref.~\onlinecite{winkler2003}.

\subsection{Zeeman interaction due to a magnetic field $\bm{B_\|}$} \label{sec:inplaneB}

To account for an in-plane magnetic field
$\vek{B}_\| = (B_x, B_y)$, we choose the asymmetric gauge
\begin{equation}
\vek{A} (z) = (z\, B_y, - z\, B_x, 0) ,
\end{equation}
so that the kinetic wave vector $\kk_\| = \bm{\mathfrak{k}}_\| +
(e/\hbar) \, \vek{A}$ becomes
\begin{equation}
\label{eq:k_operator}
\kk_\| = \left(\begin{array}{c}
\mathfrak{k}_x + \left(e/\hbar\right) z B_y \\[0.5ex]
\mathfrak{k}_y - \left(e/\hbar\right) z B_x
\end{array} \right) \, ,
\end{equation}
where $\bm{\mathfrak{k}}_\|$ denotes the canonical wave vector.
The components of $\vek{k}_\|$ thus do not commute with $k_z
\equiv -i\partial_z$.

To evaluate the Zeeman interaction linear in the field
$\vek{B}_\|$, we may restrict ourselves to the terms in $H'_k$
that are linear in $\kk_\|$.  We denote these terms by
$H^{\prime \, (1)}_k$.  As mentioned above, $H^{\prime \, (1)}_k$
generally includes symmetrized products $\{ \vek{k}_\|, k_z\}$.
Terms of higher order in $\vek{k}_\|$ or $\vek{B}_\|$ are ignored
as they do not contribute to the Zeeman interaction linear in
$\vek{B}_\|$.  To evaluate the Zeeman interaction at the subband
edge, we take $\vek{\mathfrak{k}}_\| = 0$.

If we use the proper eigenstates of $H_0$, it is sufficient to
evaluate $H' = \vek{H}'_B\cdot\vek{B}_\| + H^{\prime \, (1)}_k$
in first order perturbation theory.  By definition, these terms
completely take into account the Zeeman interaction linearly
proportional to $\vek{B}_\|$.  More explicitly, we have
\begin{equation}
\bigl[ H_Z^{(\alpha)} \bigr]_{\alpha\alpha'}
= \braket{\alpha | \vek{H}'_B\cdot\vek{B}_\| + H^{\prime \, (1)}_k
	| \alpha'},
\end{equation}
compare Eq.\ (\ref{eq:zeeman}).  Starting from $\tilde{H}_0$ as
unperturbed Hamiltonian, we need typically one more order of
perturbation theory to construct nontrivial wave functions
$\{\ket{\alpha}\}$ that incorporate band mixing.

\subsection{Zeeman interaction due to a magnetic field $\bm{B_z}$} \label{sec:outofplaneB}

The contribution to the Zeeman interaction from $H'_{B_z}$ can be
accounted for in first-order (starting from $H_0$) or second-order
(starting from $\tilde{H}_0$) perturbation theory as discussed
above.
The nontrivial contribution to the Zeeman interaction is due to
$H'_k$.  First we construct from $H'_k$ the pair of $N \times N$
operators $H'_\pm$ that contain only the prefactors for the terms
linear in $k_\pm \equiv k_x \pm ik_y$ (ignoring terms of higher
order in $k_\pm$), i.e.,
\begin{equation}
  H'_k = H'_+ k_+ + H'_- k_- + \mathcal{O}(k_\|^2) .
\end{equation}
The $N\times N$ operators $H'_\pm$ are Hermitian adjoint pairs
with $H_\mp^{\prime\,\dagger} = H'_\pm$.  The quantities $H'_\pm$
are generally proper operators in the sense that they may contain
powers of $k_z = -i\partial_z$.  Yet for a magnetic field $B_z$ we
have $[k_\pm,k_z] = 0$.

To evaluate the Zeeman interaction linear in $B_z$, we treat
$H'_+ k_+ + H'_- k_-$ as small quantity.  In second order L\"owdin
partitioning, we get
\begin{widetext}
\begin{align}
H'_{\alpha\alpha'} = & \hspace{0.9em}
\sum_\beta \frac{1}{E_\alpha - E_\beta}
\bigl(  \braket{\alpha|H'_+|\beta} \braket{\beta|H'_-|\alpha'} k_+k_-
+ \braket{\alpha|H'_-|\beta} \braket{\beta|H'_+|\alpha'} k_-k_+
\bigr)
\nonumber \\* &
+ \sum_\beta \frac{1}{E_\alpha - E_\beta}
\left(
\braket{\alpha|H'_+|\beta} \braket{\beta|H'_+|\alpha'} \, k_+^2
+ \braket{\alpha|H'_-|\beta} \braket{\beta|H'_-|\alpha'} \, k_-^2
\right) .
\end{align}
Here, the sums run over all intermediate states
$\beta \ne \alpha, \alpha'$ and we assumed
$E_\alpha = E_{\alpha'}$.
Equation (\ref{eq:k_com}) implies that in the presence of a
magnetic field $B_z$, we have
\begin{equation}
k_\pm k_\mp = k_\|^2 \mp B_z .
\end{equation}
Note also that for any pair of states $\ket{\alpha}$ and
$\ket{\beta}$ we have
$\braket{\alpha|H'_-|\beta} = \braket{\beta|H'_+| \alpha}^\ast$.
Thus we get
\begin{align}
H'_{\alpha\alpha'} = & \hspace{0.9em}
\sum_\beta \frac{1}{E_\alpha - E_\beta}
\bigl(  \braket{\alpha|H'_+|\beta} \braket{\alpha'|H'_+|\beta}^\ast
+ \braket{\beta|H'_+|\alpha}^\ast \braket{\beta|H'_+|\alpha'} \bigr)
k_\|^2
\nonumber \\* &
- \sum_\beta \frac{1}{E_\alpha - E_\beta}
\bigl(  \braket{\alpha|H'_+|\beta} \braket{\alpha'|H'_+|\beta}^\ast
- \braket{\beta|H'_+|\alpha}^\ast \braket{\beta|H'_+|\alpha'} \bigr)
B_z
\nonumber \\* &
+ \sum_\beta \frac{1}{E_\alpha - E_\beta}
\left(
\braket{\alpha|H'_+|\beta} \braket{\beta|H'_+|\alpha'} \, k_+^2
+ \braket{\beta|H'_+|\alpha}^\ast \braket{\alpha'|H'_+|\beta}^\ast \, k_-^2
\right) .
\label{eq:mg:loewdin}
\end{align}
\end{widetext}

The perturbative scheme (\ref{eq:mg:loewdin})
naturally provides a more comprehensive description of the
in-plane dynamics in a 2D system.  Similar to Eq.\
(\ref{eq:g_tensor_parts}), the in-plane motion for the
degenerate subspace $\{\ket{\alpha}, \ket{\alpha'}\}$ is
characterized by an inverse effective mass
$\mu = \mu_0 + \Delta\mu$, where $\mu_0$ denotes the contribution
to $\mu$ due to remote bands outside the $N$-dimensional space,
i.e., it stems from the terms $\propto k_\|^2$ in $H'_k$.  The
correction $\Delta\mu$ (like $\Delta \vek{g}$) appears due to the
coupling of the subspace $\{ \ket{\alpha}, \ket{\alpha'} \}$ to
the intermediate states $\{ \ket{\beta} \}$ within the
$N$-dimensional space.

The first term in Eq.\ (\ref{eq:mg:loewdin}) gives an isotropic
contribution to $\Delta\mu$ for the in-plane motion, the second
term gives $\Delta \vek{g}$.  The last term is an anisotropic correction
to the dispersion [note $k_+^2 + k_-^2 = 2(k_x^2 - k_y^2)$].  If
the set of intermediate states $\{\ket{\beta}\}$ is complete
within the $N$-dimensional space, Eq.\ (\ref{eq:mg:loewdin}) gives
exact expressions for $\Delta \mu$ and $\Delta \vek{g}$.  Note that this
refers to the limit of small wave vectors $k_\|$, i.e., neglecting
higher-order corrections to $\Delta \mu$ and $\Delta \vek{g}$ that are
themselves a function of $k_\|$ or $B_z$.

\subsection{Analytical Results} \label{sec:anaresults}

To obtain analytical expressions for the nonzero components of
$\vek{g}$, we describe the quasi-2D holes by means of the
$4\times 4$ Luttinger $H$ for the bulk valence band using the same
notation as in Ref.~\onlinecite{winkler2003}.  We assume an
infinitely deep rectangular QW of width $w$.  Using the
band-diagonal Hamiltonian $\tilde{H}_0$ as unperturbed Hamiltonian,
we obtain in second-order perturbation theory for the lowest subband
HH1
\begin{equation}
\label{eq:g_tensor}
\vek{g} =
\begin{pmatrix}
g_{xx} & 0 & g_{xz}\\
0 & g_{yy} & 0\\
g_{zx} & 0 & g_{zz}
\end{pmatrix} ,
\end{equation}
with
\begin{subequations}
\begin{align}
g_{xx} = & \left[\frac{3 \, \kappa \left(\gamma_3 - \gamma_2
	\right) \pi^2 \hbar^2}{\Delta^{hl}_{11} \, w^2} \sin^2 \theta
+ \frac{3 q}{2} (1 + \sin^2 \theta)\right]
\nonumber \\ & {} \times
\left(2 - 3 \sin^2 \theta \right) ,
\label{eq:gxx} \\
g_{zx} = & \left[ - \frac{6 \, \kappa \left(\gamma_3 - \gamma_2 \right)
	\pi^2 \hbar^2} {\Delta^{hl}_{11} \, w^2}
+ \frac{3q}{2} \right]
\left(2 - 3 \sin^2 \theta \right) \sin \theta \cos \theta ,
\label{eq:gzx} \\
g_{yy} = & \left[\frac{3 \, \kappa \left(\gamma_3 - \gamma_2
	\right) \pi^2 \hbar^2}{\Delta^{hl}_{11} \, w^2} \sin^2 \theta
- \frac{3 q}{2}\right]
\left(2 - 3 \sin^2 \theta \right) ,
\label{eq:gyy} \\
g_{xz} = & \frac{3}{2} q (2 + 3\sin^2\theta) \sin \theta \cos \theta ,
\nonumber \\ & {}
+ \frac{128 \left(\gamma_3 - \gamma_2 \right) \hbar^2}
{3 \, \Delta^{hl}_{12} \, w^2}
\bigl[ 3\gamma_2 \left(2 - \sin^2\theta \right) \sin^2\theta
\nonumber \\ & \hspace{2em}
+ \gamma_3 \left(2 - 3 \sin^2\theta + 3 \sin^4\theta \right) \bigr]
\sin \theta \cos \theta ,
\label{eq:gxz} \\
g_{zz} = & - 6\kappa - \frac{3\, q}{2} (9 - 4 \sin^2 \theta
+ 3 \sin^4\theta)
\nonumber \\ &
{} + \frac{128 \, \hbar^2}{3 \, \Delta^{hl}_{12} \, w^2}
\bigl[ 3 \gamma_2^2 \cos^2 \theta \sin^4\theta
\nonumber \\ & \hspace{2em}
+ 2 \gamma_2 \gamma_3 \sin^2 \theta \left(4 - 6 \sin^2 \theta
+ 3 \sin^4 \theta \right) \nonumber \\
& \hspace{2em}
+  \gamma_3^2 \cos^2 \theta \left(2 - 6 \sin^2 \theta
+ 3 \sin^4 \theta \right) \bigr] ,
\label{eq:gzz}
\end{align}
\label{eq:g_analytical}
\end{subequations}
where $\Delta^{hl}_{\alpha\beta} \equiv E^h_\alpha - E^l_{\beta}$ with
\begin{subequations}
\begin{align}
\Delta^{hl}_{11} & = \frac{\pi^2 \hbar^2}{2 w^2}
\bigl[ \gamma_2 \left(2 - 3 \sin^2 \theta \right)^2
+ 3 \gamma_3 \left( 4 - 3 \sin^2 \theta \right) \sin^2 \theta \bigr] , \\
\Delta^{hl}_{12} & = \frac{\pi^2 \hbar^2}{4 w^2}
\bigl[ 6 \gamma_1 + 5 \gamma_2 \left(2 - 3 \sin^2 \theta \right)^2
\nonumber \\[-1.0ex] & \hspace{4.5em}{}
+ 15 \gamma_3 \left( 4 - 3 \sin^2 \theta \right) \sin^2 \theta \bigr] .
\end{align}
\end{subequations}
For an infinitely deep rectangular QW, the components of the
tensor $\vek{g}$ are thus independent of the well width $w$.

The analytical results are presented in
Figs.~\ref{fig:wellwidth}(b)-(d), using $\gamma_1 = 6.85$,
$\gamma_2 = 2.10$, $\gamma_3 = 2.90$, $\kappa = 1.20$ and
$q = 0.01$, appropriate for GaAs. \cite{winkler2003} We discuss
the analytical results in Sec.~\ref{sec:numcalculations} where we compare
them with accurate numerical calculations.

\subsection{Numerical calculations} \label{sec:numcalculations}

To obtain accurate numerical results for the Zeeman interaction, we
describe the quasi-2D holes using the $8\times 8$ Kane Hamiltonian
as described in Ref.~\onlinecite{winkler2003}.  The
unperturbed Hamiltonian is here the full Hamiltonian $H_0$ that we
solve as discussed in Ref.~\onlinecite{winkler1993}.  Consistent
with the experiments, we consider a symmetric GaAs-AlAs QW
of well width 12~nm, using the band parameters listed in Ref.~\onlinecite{winkler2003}.  The numerical results are presented in Figs.~\ref{fig:wellwidth}(b)-(d).

We see in Eq.\ (\ref{eq:g_analytical}) that the components $g_{xz}$
and $g_{zz}$ depend on the coupling of the ground subband HH1 to the
first excited LH subband LH2.  Here, the analytical model of an
infinitely deep QW overestimates the subband gap $\Delta^{hl}_{12}$,
which becomes the most significant for narrow QWs.  This effect can
be seen clearly for $g_{zz}$ in Eq.\ (\ref{eq:gzz}) and
Fig.~\ref{fig:wellwidth}(d), where the dominant first-order term
$-6\kappa$ is negative and the second-order corrections
$\propto 1/\Delta^{hl}_{12}$ are positive. The components $g_{xx}$,
$g_{zx}$, and $g_{yy}$ depend on the coupling of the ground subband
HH1 to the lowest LH subband LH1.  Both HH1 and LH1 are strongly
confined so that the model of an infinitely deep QW is more accurate
for these components of $\vek{g}$ and the well width-dependent
deviations of the numerical calculations from the analytical model
are less pronounced.

For the special cases $\theta = 0^\circ$ (growth direction [001], group $D_{2d}$), $\theta = 54.7^\circ$ (growth direction [111], group $D_{3d}$), and  $\theta = 90^\circ$ (growth direction [110], group $C_{2v}$) both the analytical and the numerical results agree with the findings in Table~\ref{tab:invar} based on the theory of invariants.

\section{Experiments}\label{sec:experiments}

\subsection{Sample design}\label{sec:sampledesign}

The samples were grown by molecular beam epitaxy on undoped GaAs substrates with different growth directions (sample A: [113]A, B: quasi-[111]B, C: [110]), following the design shown in Ref.~\onlinecite{gradl2014}. Note that sample B was not exactly grown along the [111]B direction. Due to a significantly cleaner growth process, a sligthly tilted substrate ($2.8^\circ$ tilt towards the [$\bar{1}\bar{1}2$] direction) was used \cite{vina1986, tsutsui1990, herzog2012}. This led to an effective growth direction of about [$\overline{10}\:\overline{10}\:\overline{9}$]. For convenience, the growth direction of sample B will be denoted by quasi-[111].

First, a highly $n$-doped GaAs layer is grown, serving as a conductive backgate which is contacted after the growth process from the top by alloying indium contacts. After a separating superlattice, the active region of the sample is stacked on top, which consists of two nominally undoped GaAs QWs with 5~nm and 12~nm width, respectively, embedded in AlAs barrier material. The QWs are separated by an 8~nm thin AlAs layer, allowing electron tunneling between the QWs. The top gate is realized by a 7~nm thin NiCr layer, stacked on a 15~nm thick SiO$_2$ layer, which are both thermally evaporated on top of the sample. Only the 12~nm wide QWs are investigated in this study.

\subsection{Experimental methods} \label{sec:expmethods}

For the photoluminescence (PL) and TRKR measurements, a pulsed Ti-Sapphire laser system is used as a light source. The system operates with a repetition rate of 80~MHz, resulting in a time delay of 12.5~ns between subsequent pulses, which have a length of 2~ps, corresponding to a spectral width of 1~meV. A beam splitter divides the laser pulses into a pump and probe beam.

In the PL experiments, only the pump beam is focused to a diameter of about 80 $\mu$m on the sample with an achromatic lens. The excitation density of about 5~W~cm$^{-2}$ leads to an optically induced carrier density of about $1 \times 10^{9}$~cm$^{-2}$. The emitted PL is collected by the same lens and guided into a spectrometer with a Peltier-cooled CCD chip.

In the TRKR experiments, the pump beam is circularly polarized and, depending on the helicity, generates electron-hole pairs in the QW with spins aligned either parallel or antiparallel to the beam direction perpendicular to the plane of the QW. Here, we selectively excite and probe the 12~nm wide QWs by tuning the laser into resonance with the transition energy of the wide QW, so that creation of electron-hole pairs in the narrow QW can be neglected. The linearly polarized probe beam passes a mechanical delay line, which provides a variable time delay between pump and probe pulses.  Then it is focused on the same spot on the sample as the pump beam. The axis of linear polarization of the probe beam is rotated by a small angle because of the magneto-optical Kerr effect due to the injected spin polarization. Note that the Kerr effect is only sensitive to an out-of-plane spin polarization in this geometry. The small rotational angle of the linear polarization is measured on an optical bridge and a lock-in scheme is used to increase sensitivity.

The measurements are performed in an optical cryostat with a $^3$He insert, providing sample temperatures of about 1.2~K. Magnetic fields of up to 11.5~T can be applied via superconducting  coils. The samples are mounted on a piezoelectric rotary stepper, enabling rotations about the out-of-plane direction of the samples. An optical feedback via a camera system is used to adjust the angle of the sample with respect to the magnetic field direction. Additionally, the sample plane can be tilted with respect to the magnetic field direction, allowing also out-of-plane magnetic field components. During the TRKR measurements, typically a field of 1~T is applied.

\subsection{Creating a 2D hole system} \label{sec:holecreation}

\begin{figure}[t]
\begin{center}\includegraphics[width=8.5cm]{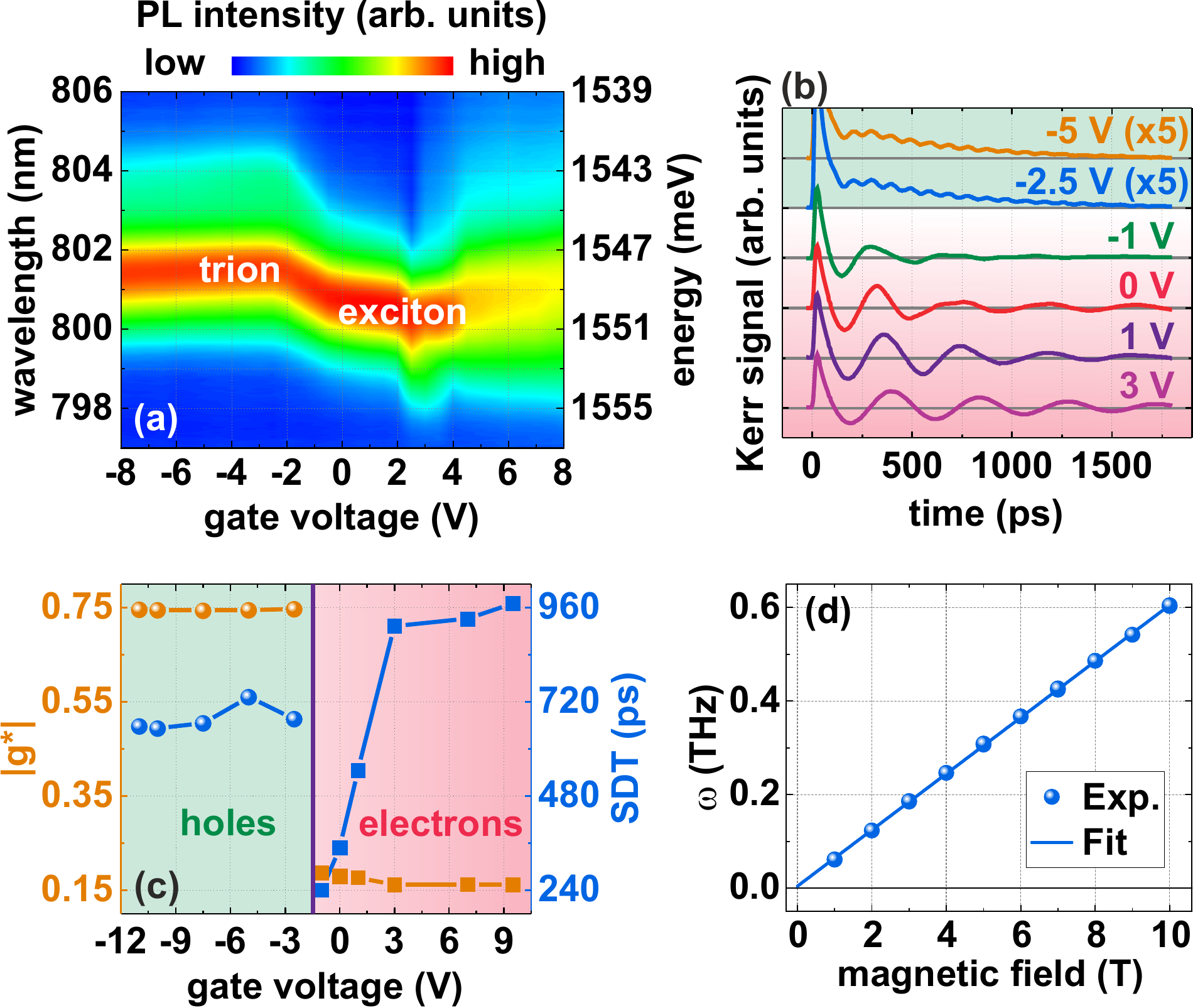}\end{center}
\caption{(a) Gate-dependent PL measurements on sample A. (b) Gate-dependent Kerr traces measured on sample A with an in-plane magnetic field $B~=~1~\mathrm{T}$, applied in the [33$\bar{2}$] direction. (c) $g$ factors and spin dephasing times (SDT) extracted from the measurements shown in (b). (d) Larmor precession frequency ($\omega$) extracted from Kerr traces as a function of the magnetic field applied in-plane in the [33$\bar{2}$] direction for a gate voltage of $-2.5$~V. The effective $g$ factor of about 0.75 is determined via a linear fit. \label{fig:gatedependence}}
\end{figure}

The optical creation of a 2D hole system is crucial to be able to observe hole-spin coherence. Our approach is described in the following on the basis of sample A.

First, we check the functionality of our applied gate with PL measurements on the 12~nm wide QW, shown in Fig.~\ref{fig:gatedependence}(a). Here, we excite the sample with a laser wavelength of 780~nm, i.e., an energy of about 1591~meV. The emitted PL light shows a clear dependence on the applied gate voltage. At around 2~V, the emission takes place at the highest energy which indicates the point where the gate voltage compensates internal electric fields of the sample so that the potential of the QW along the growth axis is flat. For higher gate voltages a bleaching can be seen, which can be attributed to a separation of the carriers in the QW along the growth axis and therefore a suppressed recombination. Towards lower gate voltages the emission shifts to higher wavelengths, showing the transition from neutral excitons to trions which are most prominent at around $-2$~V. Considering the TRKR measurements shown in Fig.~\ref{fig:gatedependence}(b), we can assign a positive charge to the trions, which will be discussed in detail in the following.

The Kerr traces show a pronounced peak around zero time delay between pump and probe pulse due to the injection of the out-of-plane spin polarization. This is followed by partial recombination of the electron-hole pairs on a timescale of about 100~ps. After that, an exponentially damped oscillation can be seen in the Kerr signal. The oscillation is attributed to the Larmor precession of the spin-polarized carriers in the QW, while the damping indicates the ensemble spin dephasing. Comparing the two top traces to the bottom traces, two clear differences are observable. First, a large discrepancy in the precession frequency becomes visible. Second, the two top traces show an additional, non-oscillatory and exponentially damped component of the Kerr signal that represents a non-precessing component of the spin dynamics. This feature plays a crucial role in our determination of the tensor
$\vek{g}$. Its origin will be discussed in detail in Sec.~\ref{sec:determineholegtensor}.

\begin{figure}[t]
\begin{center}\includegraphics[width=8.5cm]{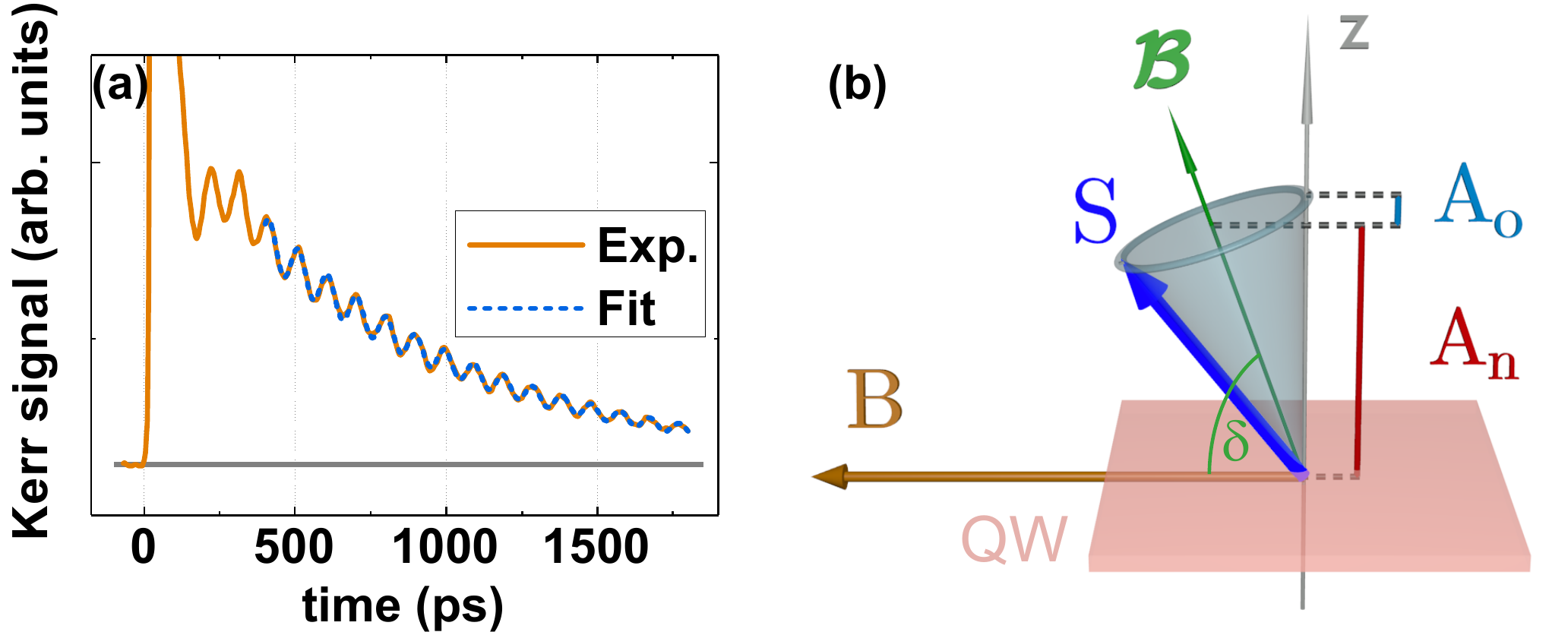}\end{center}
\caption{(a) Exemplary fit of a Kerr trace based on Eq.\ (\ref{eq:trkrfitformula}). (b) Schematic picture for a tilted effective field $\vekc{B}$ pointing out of the QW plane for an in-plane magnetic field $\vek{B}$ and the resulting precession of the spins $\vek{S}$. Animations are available online. \label{fig:precessionaxis}}
\end{figure}

We fit the data by an exponentially damped cosine (prefactor $A_\mathrm{o}$) combined with an exponentially damped term (prefactor $A_\mathrm{n}$) for the non-oscillatory component:
\begin{equation}
A(t) = A_\mathrm{o}\ \mathrm{e}^{-t/\tau_\mathrm{o}}\cos(\omega t) +  A_\mathrm{n}\ e^{-t/\tau_\mathrm{n}} .
\label{eq:trkrfitformula}
\end{equation}
An exemplary fit is depicted in Fig.~\ref{fig:precessionaxis}(a). Hence these fits yield both the precession frequency $\omega$ (which is directly related to the effective $g$ factor via $g^\ast = \hbar\omega/\mu_B B$) and the spin dephasing time  $\tau_\mathrm{o}$\cite{foot:taun}. The extracted values are presented in Fig.~\ref{fig:gatedependence}(c). Note that we can only determine the absolute value of $g^\ast$. The $g$ factor above gate voltages of $-2$~V is about 0.16. This value can be attributed to electron-spin precession, considering the transition energy of the GaAs/AlAs QW of about 1550~meV in comparison to a study of Yugova et al. \cite{yugova2007} They found an electron $g$ factor of about 0.20 for a transition energy of about 1555~meV in GaAs/Al$_x$Ga$_{1-x}$As QWs for different Al concentrations, which is in reasonable agreement with our measurements. A significantly higher $g$ factor is obtained for negative gate voltages. Figure~\ref{fig:gatedependence}(d) exemplarily shows the magnetic field dependence of the precession frequency for a gate voltage of $-2.5$~V. A linear fit yields an effective $g$ factor of about 0.75 indicating hole spin precession. In a previous work on a similar sample we could already measure a hole $g$ factor of about 0.7 in this in-plane crystallographic direction \cite{gradl2014}.

The switch between hole- and electron-dominated regimes in the wide QW can be understood in the following way: For a negative gate voltage the electrons tunnel out of the QW towards the top contact while the holes remain in the QW. For a positive gate voltage, on the other hand, electrons from the top contact tunnel into the QW and create an excess of electrons.

The spin dephasing time of the holes shown in Fig.~\ref{fig:gatedependence}(c) stays relatively constant at a value of about 700~ps in the gate-voltage range where we observe long-lived hole spin precession. This observation, combined with the fact that the amplitude of the TRKR hole signal does not show a strong dependence on gate voltage in this range, indicates that there is no significant tuning of the hole density by the gate voltage. Previously, a pronounced decrease of hole spin dephasing time with increasing hole density was observed by several groups \cite{BAYLAC199557, syperek2007, kugler2011}. By contrast, the electron spin dephasing time is most likely limited by the carrier lifetime in the QW. A closer look at the Kerr traces [Fig.~\ref{fig:gatedependence}(b)] in this gate voltage range shows a small beating of a second precession frequency which can be attributed to an increasing hole density towards lower gate voltages. At high gate voltages the spin dephasing time of the electrons stays relatively constant around 900 ps. Here, the electrons are not affected by the holes anymore as these quickly recombine, with the electrons tunneled into the QW.

The focus of this work is the study of hole-spin dynamics, hence all subsequent TRKR measurements on sample A were carried out using a gate voltage of $-2.5$~V. For samples B and C, a study of gate-voltage-dependent TRKR measurements (not shown) revealed that the hole-dominated regime is realized at a gate voltage of 0~V, so this voltage was used for subsequent measurements on these samples.

\subsection{Determination of the tensor $\vek{g}$} \label{sec:determineholegtensor}

We start with a general discussion of observable effects due to a
Zeeman term (\ref{eq:zeeman}) and how these effects allow one to
determine $\vek{g}$.  Given the tensor $\vek{g}$, which is in
general neither symmetric nor antisymmetric, the Zeeman splitting of
the energy levels in an external magnetic field $\vek{B}$ becomes \cite{abragam1970}
\begin{equation}
\label{eq:zeeman_split}
\Delta E = \mu_\mathrm{B} |\vek{g} \cdot \vek{B}|
= \mu_\mathrm{B} \sqrt{\vek{B}^\dagger \cdot \vek{G} \cdot \vek{B}} ,
\end{equation}
where
\begin{equation}
\vek{G} \equiv \vek{g}^\dagger \cdot \vek{g}
\label{eq:Gdefinition}
\end{equation}
is a symmetric, positive definite second-rank tensor.  Here we use the symmetry analysis in Sec.~\ref{sec:sym} to reduce the number of independent parameters considered for $\vek{G}$.  Equation (\ref{eq:g_tensor}) gives
\begin{equation}
\label{eq:Gtensor}
\vek{G} =
\begin{pmatrix}
g_{xx}^2 + g_{zx}^2 & 0 & g_{xx} g_{xz} + g_{zx} g_{zz}\\
0 & g_{yy}^2 & 0\\
g_{xx} g_{xz} + g_{zx} g_{zz} & 0 & g_{xz}^2 + g_{zz}^2
\end{pmatrix} .
\end{equation}
Fitting the Kerr traces by Eq.\ (\ref{eq:trkrfitformula})
yields the Larmor precession frequency $\omega$ that determines the
Zeeman splitting
\begin{equation}
  \hbar\omega
  = \mu_\mathrm{B} \sqrt{\vek{B}^\dagger \cdot \vek{G} \cdot \vek{B}} .
\end{equation}
Using the parameterization $\vek{B} = B_0 (\cos\alpha \sin\beta, \allowbreak \sin\alpha \sin\beta, \allowbreak \cos\beta)$ [see Fig.~\ref{fig:113_tensor}(a)], we get
\begin{subequations}
\label{fitformula}
\begin{align}
g^\ast (\alpha,\beta) = & \frac{\hbar\omega}{\mu_BB_0} \\
 = & \bigl[ G_{xx} \cos^2\alpha\sin^2\beta + G_{yy}\sin^2\alpha \sin^2\beta
  \nonumber \\
 & {} + 2G_{xz}\cos\alpha\sin\beta\cos\beta + G_{zz}\cos^2\beta
\bigr]^{1/2} .
\end{align}
\end{subequations}
A measurement of $g^\ast (\alpha,\beta)$ thus allows us to determine the tensor $\vek{G}$ characterizing the Zeeman splitting of the energy levels \cite{abragam1970}.

The symmetric tensor $\vek{G}$ has four independent components.  The tensor $\vek{g}$, on the other hand,  has five independent parameters according to  Eq.\ (\ref{eq:g_tensor}).  (In the most general case the symmetric tensor $\vek{G}$ has six independent parameters, whereas $\vek{g}$ has nine independent parameters.)
The presence of additional independent parameters characterizing
$\vek{g}$ as compared to $\vek{G}$ is related to the fact that,
according to Eq.\ (\ref{eq:zeeman}), the tensor $\vek{g}$ describes
the orientation of the spin $\vek{S}$ in the external field
$\vek{B}$.  Here, $\vek{g}$ defines a similarity transformation
\begin{equation}
\vekc{B} = \vek{g} \cdot \vek{B} ,
\label{eq:definitionBeff}
\end{equation}
so that in an eigenstate of the Zeeman Hamiltonian
(\ref{eq:zeeman}), the spin $\vek{S}$ gets aligned parallel or
antiparallel to the effective field $\vekc{B}$.  A noneigenstate, on
the other hand, precesses about the effective field $\vekc{B}$.
While the magnitude of $\vekc{B}$ determines the precession
frequency $\omega$ (and thus it determines $\vek{G}$), the direction
of $\vekc{B}$ represents the precession axis, i.e., the spins
precess on a cone about $\vekc{B}$, as shown in
Fig.~\ref{fig:precessionaxis}(b).  A tensor $\vek{g}$ that is not just proportional to a scalar implies that the precession axis $\vekc{B}$ is, in general, not parallel to the external field $\vek{B}$.  A measurement of the direction of $\vekc{B}$ as a function of the direction of the external field
$\vek{B}$ thus allows one to determine the remaining independent
parameters that characterize $\vek{g}$ as compared with $\vek{G}$.

Here the key feature of our experiments allowing us to access this
information lies in the fact that the Kerr effect is only sensitive
to an out-of-plane spin polarization, i.e., the projection of the
spin polarization on the $z$ axis.  Therefore, the shape of the TRKR
signal changes for a precession axis $\vekc{B}$ that is tilted out
of the QW plane, as expressed in Eq.\ (\ref{eq:trkrfitformula}) by
the presence of the non-oscillatory component $\propto A_\mathrm{n}$.  It is
this aspect of our experiments that allows us to determine not only
the magnitude but also the direction of $\vekc{B}$, which, in turn,
yields the full tensor $\vek{g}$.

To relate the experimental information contained in the amplitude
$A_\mathrm{n}$ with the remaining parameters characterizing $\vek{g}$ (for
given $\vek{G}$), we proceed in several steps.
First, we note that the symmetric Zeeman tensor $\vek{G}$ can be
brought to diagonal form by means of an orthogonal
transformation~$\vekc{O}$
\begin{equation}
\vekc{G} = \vekc{O} \cdot \vek{G} \cdot \vekc{O}^{-1} ,
\end{equation}
where the columns of $\vekc{O}^{-1}$ are the principal axes of $\vek{G}$.  We define
\begin{equation}
\tilde{\vek{g}} = \vekc{O}^{-1} \cdot \sqrt{\vekc{G}} \cdot \vekc{O} .
\end{equation}
By definition, the tensor $\tilde{\vek{g}}$ is symmetric ($\tilde{g}^\dagger = \tilde{g}$) and we have
\begin{equation}
  \label{eq:gtilde:from:G}
  \vek{G} = \tilde{\vek{g}}^\dagger \cdot \tilde{\vek{g}} ,
\end{equation}
i.e., $\tilde{\vek{g}}$ and $\vek{g}$ describe the same Zeeman splitting (\ref{eq:zeeman_split}), though in general the symmetric tensor $\tilde{\vek{g}}$ (like $\vek{G}$) depends on fewer independent parameters than the asymmetric tensor~$\vek{g}$.

Indeed, we have
\begin{equation}
\vek{g} = \vek{d} \cdot \tilde{\vek{g}} ,
\label{eq:g_via_gtilde}
\end{equation}
where the matrix $\vek{d}$ represents a (proper or improper)
rotation that affects the alignment of the spin relative to the
field $\vek{B}$, but does not affect the Zeeman splitting
(\ref{eq:zeeman_split}). Equation (\ref{eq:zeeman}) with
$\vek{g} = \tilde{\vek{g}}$ would imply that given a magnetic field
$\vek{B}$ oriented along one of the principal axes of
$\tilde{\vek{g}}$, the spin gets oriented (anti-) parallel to
$\vek{B}$.  The matrix $\vek{d}$ can be parameterized by, e.g., up
to three Euler angles.  The independent components of
$\tilde{\vek{g}}$ (or $\vek{G}$) together with the Euler angles
characterizing $\vek{d}$ thus provide a parameterization of
$\vek{g}$ (apart from a sign ambiguity discussed below).

In the present experiments, inverting Eq.\ (\ref{eq:gtilde:from:G}) yields (apart from a sign)
\begin{equation}
    \tilde{g}_{yy} = \sqrt{G_{yy}}
\end{equation}
and (apart from another overall sign)
\begin{subequations}
  \label{eq:gtilde_components}
  \begin{align}
  \tilde{g}^\pm_{xx} = &
  \frac{2 G_{xz}^2 + (G_{xx} - G_{zz})
  (G_{xx} \pm \sqrt{G_{xx} G_{zz} - G_{xz}^2})}{\Gamma_\pm^2\Gamma_\mp} , \\
  \tilde{g}^\pm_{xz} = & \frac{G_{xz}}{\Gamma_\mp} , \\
  \tilde{g}^\pm_{zz} = &
  \frac{2 G_{xz}^2 + (G_{zz} - G_{xx})
  (G_{zz} \pm \sqrt{G_{xx} G_{zz} - G_{xz}^2})}{\Gamma_\pm^2\Gamma_\mp} ,
\end{align}
where
\begin{equation}
  \Gamma_\pm = \sqrt{G_{xx} + G_{zz} \pm 2 \sqrt{G_{xx} G_{zz} - G_{xz}^2}} .
\end{equation}
\end{subequations}
While the components $\tilde{g}^+_{ij}$ and $\tilde{g}^-_{ij}$ may
differ, in general, both in sign and magnitude, they define via Eq.\
(\ref{eq:g_via_gtilde}) the same tensor $\vek{g}$.  To show this,
we define from Eq.\ (\ref{eq:gtilde_components}) a reduced tensor
\begin{equation}
\label{eq:red:gtilde}
\tilde{\vek{g}}_\mathrm{red}^\pm=
\begin{pmatrix}
\tilde{g}^\pm_{xx} & \tilde{g}^\pm_{xz}\\
\tilde{g}^\pm_{xz} & \tilde{g}^\pm_{zz}
\end{pmatrix} ,
\end{equation}
where
$\tilde{\vek{g}}_\mathrm{red}^+ = \tilde{\vek{d}}_\mathrm{red} \cdot
\tilde{\vek{g}}_\mathrm{red}^-$
with an orthogonal $2\times 2$ matrix $\tilde{\vek{d}}_\mathrm{red}$ with
$\det \tilde{\vek{d}}_\mathrm{red} = -1$, i.e., $\tilde{\vek{d}}_\mathrm{red}$
describes an improper rotation about the $y$ axis.
Similar to Eq.\ (\ref{eq:red:gtilde}), we can define the reduced tensor
\begin{equation}
\label{eq:red:g}
\vek{g}_\mathrm{red} =
\begin{pmatrix}
g_{xx} & g_{xz}\\ g_{zx} & g_{zz}
\end{pmatrix} ,
\end{equation}
so that similar to Eq.\ (\ref{eq:g_via_gtilde}) we have
$\vek{g}_\mathrm{red} = \vek{d}^+_\mathrm{red} \cdot
\tilde{\vek{g}}^+_\mathrm{red} = \vek{d}^-_\mathrm{red} \cdot
\tilde{\vek{g}}^-_\mathrm{red}$,
where $\vek{d}^+_\mathrm{red}$ and
$\vek{d}^-_\mathrm{red} = \vek{d}^+_\mathrm{red} \cdot
\tilde{\vek{d}}_\mathrm{red}$
are (proper or improper) rotations about the $y$ axis characterized
by a single angle $\gamma^\pm$.  Thus, we may work with either
$\tilde{\vek{g}}^+_\mathrm{red}$
or~$\tilde{\vek{g}}^-_\mathrm{red}$.

Since we cannot extract the sign of $g^\ast (\alpha,\beta)$ out of
the TRKR measurements we are not able to distinguish experimentally
between $\det \vek{g}_\mathrm{red} > 0$ or
$\det \vek{g}_\mathrm{red} < 0$ and $g_{yy} > 0$ or $g_{yy} < 0$.
To proceed, we thus adopt these signs from the theoretical
calculations. Since $g_{yy}= \pm \tilde{g}_{yy}$ is decoupled from the
other nonzero components of $\vek{g}$, we can directly take the sign
of $g_{yy}$ from the theoretical predictions.  The calculations also
predict $\det \vek{g}_\mathrm{red} \lessgtr 0$ for
$\theta \lessgtr 54.7^\circ$, where $\theta = 54.7^\circ$ represents
$z \parallel [111]$.  If we choose to work with $\sigma = +$ or $-$
giving
$\det \tilde{\vek{g}}^\sigma_\mathrm{red} (\theta) = \det
\vek{g}_\mathrm{red} (\theta)$,
we thus need proper rotations $\vek{d}^\sigma_\mathrm{red} (\theta) \equiv \vek{d}_\mathrm{red}$
for all angles $\theta$.  The rotation matrix $\vek{d}_\mathrm{red}$ is characterized by one angle that we denote by $\gamma$.  We determine $\gamma$ as follows.

Due to the off-diagonal component $g_{zx}\neq0$ the precession axis $\vekc{B}$ is tilted out of the QW plane even for purely in-plane external magnetic fields $\vek{B}$ (i.e., $\beta=90^\circ$):
\begin{equation}
\vekc{B} = \vek{g} \cdot \vek{B} = B_0
\begin{pmatrix}
g_{xx}\cos\alpha\\
g_{yy}\sin\alpha\\
g_{zx}\cos\alpha
\end{pmatrix} .
\label{eq:Beff_inplaneB}
\end{equation}
We define the tilt angle $\delta$ such that $90^\circ-\delta$ is the
angle between the (positive) $z$ axis and $\vekc{B}$ [see Fig.~\ref{fig:precessionaxis}(b)].  For a given
external magnetic field $\vek{B}$, the angle $\delta$ thus becomes
\begin{equation}
\tan\left[\delta(\alpha)\right] = \frac{g_{zx}\cos\alpha}{\sqrt{(g_{xx}\cos\alpha)^2+(g_{yy}\sin\alpha)^2}} .
\label{eq:deltatheo}
\end{equation}
For $\alpha = 0$, we have therefore $\tan\delta = g_{zx} / g_{xx}$, i.e.,
$\delta$ corresponds to the polar angle of the vector
$\vek{q} \equiv (g_{xx}, g_{zx})$ in the $(g_{xx}, g_{zx})$ plane.
On the other hand, as discussed above, a tilt angle
$\delta\neq0^\circ$ results in a non-oscillatory component
$A_\mathrm{n} \neq 0$ in the Kerr signal [see Eq.\
(\ref{eq:trkrfitformula})], because the spin polarization and
$\vekc{B}$ are not perpendicular to each other. Thus, the tilt angle
can be experimentally quantified [see Fig.~\ref{fig:precessionaxis}(b)]:
\begin{equation}
\left|\tan\left[\delta(\alpha)\right]\right| = \sqrt{\frac{A_\mathrm{n}(\alpha)}{A_\mathrm{o}(\alpha)}} .
\end{equation}
The modulus in this equation implies a four-fold ambiguity in the experimental determination of $\delta$.  Here, we can overcome this ambiguity by comparing with the theoretical calculations.

Finally, Eq.\ (\ref{eq:g_via_gtilde}) implies
\begin{equation}
\vekc{B}=\vek{d}\cdot\tilde{\vekc{B}} ,
\end{equation}
where $\tilde{\vekc{B}}=\tilde{\vek{g}}\cdot\vek{B}$ is the effective magnetic field which would act on the spin based on $\tilde{\vek{g}}$. Thus, $\vek{d}$ represents also the rotation that maps the effective magnetic field $\tilde{\vekc{B}}$ on $\vekc{B}$. This means for $\alpha=0^\circ$, where $\vekc{B}$ and $\tilde{\vekc{B}}$ are in the $xz$ plane, the angle $\gamma$ characterizing $\vek{d}_\mathrm{red}$ is the angle between $\vekc{B}$ and $\tilde{\vekc{B}}$:
\begin{equation}
\gamma = \delta(0^\circ) - \tilde{\delta}(0^\circ) ,
\end{equation}
where $\tan \tilde{\delta} = \tilde{g}_{zx} / \tilde{g}_{xx}$
represents the tilt of the precession axis due to $\tilde{\vek{g}}$
discussed above. Thus, we are able to determine $\vek{g}$ based on
Eq. (\ref{eq:g_via_gtilde}).

\subsection{Tensor $\vek{g}$ in [113]-grown QWs} \label{sec:holegtensor113}

\begin{figure}[t]
\begin{center}\includegraphics[width=8.5cm]{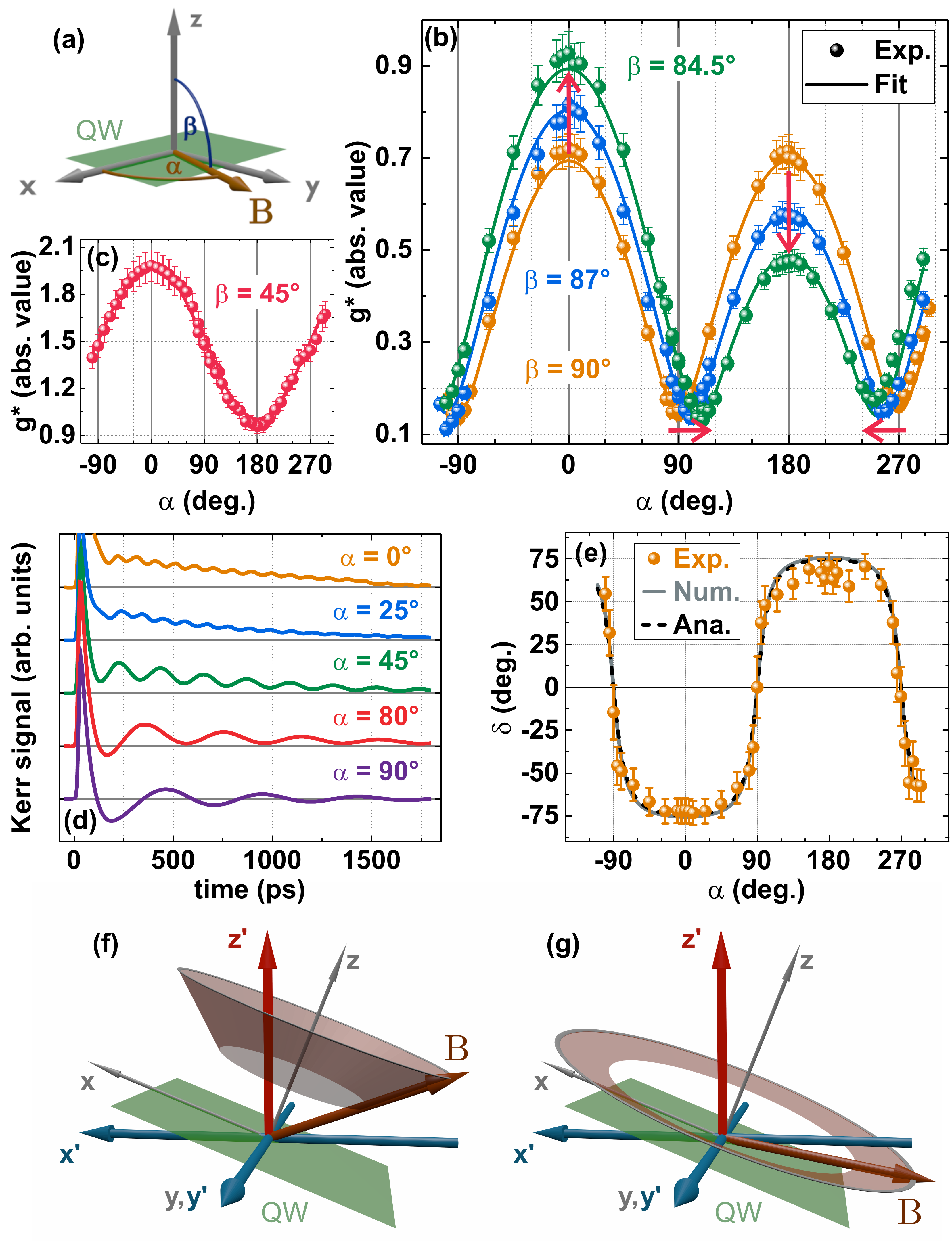}\end{center}
\caption{(a) Schematic picture of the magnetic field direction as a function of $\alpha$ and $\beta$ with respect to the crystallographic directions of the sample. (b), (c) Effective $g$ factors $g^\ast(\alpha,\beta)$ gained from TRKR measurements on sample A for different $\beta$ and a complete rotation in $\alpha$. For the fits Eq.\ (\ref{fitformula}) was used. (d) Kerr traces measured on sample A for different $\alpha$ and $\beta=90^\circ$. (e) Tilt angle $\delta$ of $\vekc{B}$ out of the QW plane extracted from TRKR measurements compared to the theoretical expectations based on the calculations shown in Secs. \ref{sec:anaresults} and \ref{sec:numcalculations}. (f), (g) Demonstrative picture for the non-diagonal $g$ tensor with defining axes ($x', y', z'$), the coordinate system of the sample ($x,y,z$) and the rotation of the magnetic field $B$ for $\beta < \beta_0 / \beta > \beta_0$. Animations are available online.}
\label{fig:113_tensor}
\end{figure}

We performed several TRKR scans of the angle $\alpha$ for different values of $\beta$ on sample A and extracted the effective $g$ factor $g^\ast(\alpha,\beta)$. Here, $\alpha = 0^\circ$, i.e., $\vek{B}$ along the $x$ axis, and $\alpha = 90^\circ$, i.e., $\vek{B}$ along the $y$ axis, correspond to the [33$\bar{2}$] direction and the [$\bar{1}$10] direction, respectively. The measurements near the in-plane direction depicted in Fig.~\ref{fig:113_tensor}(b) show two maxima of $g^\ast$ at $\alpha = 0^\circ$ and $\alpha = 180^\circ$, i.e., in $x$ and $-x$ direction, and minima around $\alpha = 90^\circ$ and $\alpha = 270^\circ$, i.e., in $y$ and $-y$ direction. This reflects the in-plane $g$ factor anisotropy, discussed in our previous work \cite{gradl2014}. An asymmetric behaviour of $g^\ast$ for $\beta < 90^\circ$ can be seen, highlighted by the red arrows.  First, $g^\ast$ at $\alpha = 0^\circ$ increases, while it decreases at $\alpha = 180^\circ$. Second, the minima, which stay at a constant value of about 0.15, shift away from $\alpha = 90^\circ$ and $\alpha = 270^\circ$, respectively, towards $\alpha = 180^\circ$. These special characteristics can be attributed to off-diagonal components of $\vek{g}$.

Applying a magnetic field with a greater out-of-plane component, i.e., $\beta = 45^\circ$, shown in Fig.~\ref{fig:113_tensor}(c), reveals a completely different behaviour of $g^\ast$. Here, only one maximum at $\alpha = 0^\circ$ and one minimum at $\alpha = 180^\circ$ emerge with considerably higher values of $g^\ast$ between about 1 and 2. This indicates a relatively large out-of-plane component $G_{zz}$. The transition from a behaviour with two maxima for nearly in-plane magnetic fields to a behaviour with only one maximum for out-of-plane magnetic fields will be discussed after the following quantitative analysis.

Using Eq.\ (\ref{fitformula}) we are able to fit the data shown in Figs.~\ref{fig:113_tensor}(b) and (c), giving $G_{xx} = 0.482\pm0.008$, $G_{zz} = 4.48\pm0.08$, $G_{xz} = 1.47\pm0.03$ and $G_{yy} = 0.0219\pm0.0003$. Choosing the signs $\det \tilde{\vek{g}}_\mathrm{red} < 0$ and $\tilde{g}_{yy} >0$ as discussed above we get
\begin{equation}
 \tilde{\vek{g}} =
 \begin{pmatrix}
 	0.213 & 0 & 0.660\\
 	0 & 0.148 & 0\\
 	0.660 & 0 & 2.01
 \end{pmatrix}, \quad \tilde{\delta} (0^\circ)=72.1^\circ .
\end{equation}
The tilt angle $\delta$ of $\vekc{B}$ is determined from in-plane field TRKR measurements [depicted in Fig.\ \ref{fig:113_tensor}(d)] which show a clear dependence of the non-oscillatory component on the magnetic field direction. For $\alpha=0^\circ$, i.e., $\vek{B}$ in the $x$ direction, the non-oscillatory component $A_\mathrm{n}$ is large, while it vanishes for $\alpha=90^\circ$, i.e., $\vek{B}$ in the $y$ direction. This indicates $g_{zx} \neq 0$ and $g_{zy}=0$ in good agreement with our theoretical predictions. By fitting the data we can extract $\delta$, which is shown in Figure \ref{fig:113_tensor}(e) compared to our theoretical expectations. Note that the sign for the experimental data is adapted to the theoretically calculated sign. A very good agreement between measurements and theory can be seen. As expected, $\delta$ is minimum in $y$ direction with $\delta(\pm90^\circ)\approx0^\circ$ and maximum in $x$ direction with $\delta(0^\circ)=-\delta(180^\circ)=-70^\circ\pm5^\circ$. This means that for a magnetic field applied parallel to the $x$ axis $\vekc{B}$ is almost perpendicular to $\vek{B}$. Note that the error margin for $\delta$ is relatively high compared to $g^\ast$ (especially for $\delta\approx0^\circ$). This is due to the fact that even tiny drifts of the Kerr signal (e.g., due to temperature fluctuations or laser stability, etc.) are affecting $A_\mathrm{n}$ while $\omega$ can be determined very accurately even for noisy experimental data. Based on $\tilde{\vek{g}}$ and $\gamma = -142^\circ$ we get using Eq.\ (\ref{eq:g_via_gtilde})
\begin{equation}
  \vek{g} =
  \begin{pmatrix}
  0.24 & 0 & 0.71\\
  0 & -0.148 & 0\\
  -0.65 & 0 & -2.0
  \end{pmatrix} .
\end{equation}
Except for $g_{yy}$, which can be determined very accurately directly via $g^\ast$, for the remaining four components a relative error margin of at least $\pm7\%$ has to be considered due to the experimental inaccuracy in $\delta(0^\circ)$.

The previously mentioned transition from a behaviour with two maxima of $g^\ast$ as a function of $\alpha$ for nearly in-plane magnetic fields ($\beta$ close to $90^\circ$) to a behaviour with only one maximum for stronger out-of-plane magnetic fields (small angles $\beta$) can be understood if we consider a simple qualitative picture of the non-diagonal tensor $\tilde{\vek{g}}$. We note that only the principal axis $y'$ of the tensor $\tilde{\vek{g}}$ agrees with the crystallographic $y$ direction $[\bar{1}10]$, whereas the remaining principal axes $x'$ and $z'$ of $\tilde{\vek{g}}$ are rotated about the $y$ axis, away from the crystallographic $x$ and $z$ axes.  As the rotation of $\vek{B}$ in our TRKR scans (parameterized by $\alpha$) is about the $z$ axis, the angle between $\vek{B}$ and the $z'$ axis changes as a function of $\alpha$. This angle is an important parameter for the measured effective $g$ factor $g^\ast (\alpha,\beta)$, since the out-of-plane components of $\vek{G}$ and therefore the $g$ factor in $z'$ direction dominate ($G_{zz} \gg G_{xx}, G_{yy}$). This leads to two distinct regimes. If the angle between the $z'$ axis and $\vek{B}$ is below $90^\circ$ for a complete scan $0 \le \alpha \le 360^\circ$, only one maximum emerges for $\alpha = 0^\circ$, where the direction of $\vek{B}$ is close to the $z'$ axis, while for $\alpha = 180^\circ$ a minimum arises, as illustrated in Fig.~\ref{fig:113_tensor}(f). Otherwise, two maxima emerge for $\alpha = 0^\circ$ and $\alpha = 180^\circ$, where the direction of $\vek{B}$ is close to the $z'$ axis, and two minima arise when $\vek{B}$ is perpendicular to the $z'$ axis, as illustrated in Fig.~\ref{fig:113_tensor}(g). Animations for both cases are available online. The tilt angle can be calculated, e.g., from the derivative of Eq.\ (\ref{fitformula}) to determine the extrema of $g^\ast$. This shows that the transition from one regime to the other occurs when
\begin{equation}
\beta_0 = \arctan\left(\frac{G_{xz}}{G_{xx}-G_{yy}}\right) .
\end{equation}
For the [113]-grown QW we get $\beta_{0, [113]} = 72.6^\circ$. This implies that the $z'$ axis is tilted by $90^\circ - \beta_{0, [113]} = 17.4^\circ$ away from the $z$ axis towards the $x$ axis.

\subsection{Tensor $\vek{g}$ in quasi-[111]-grown QWs} \label{sec:holegtensor111}

\begin{figure}[t]
\begin{center}\includegraphics[width=8.5cm]{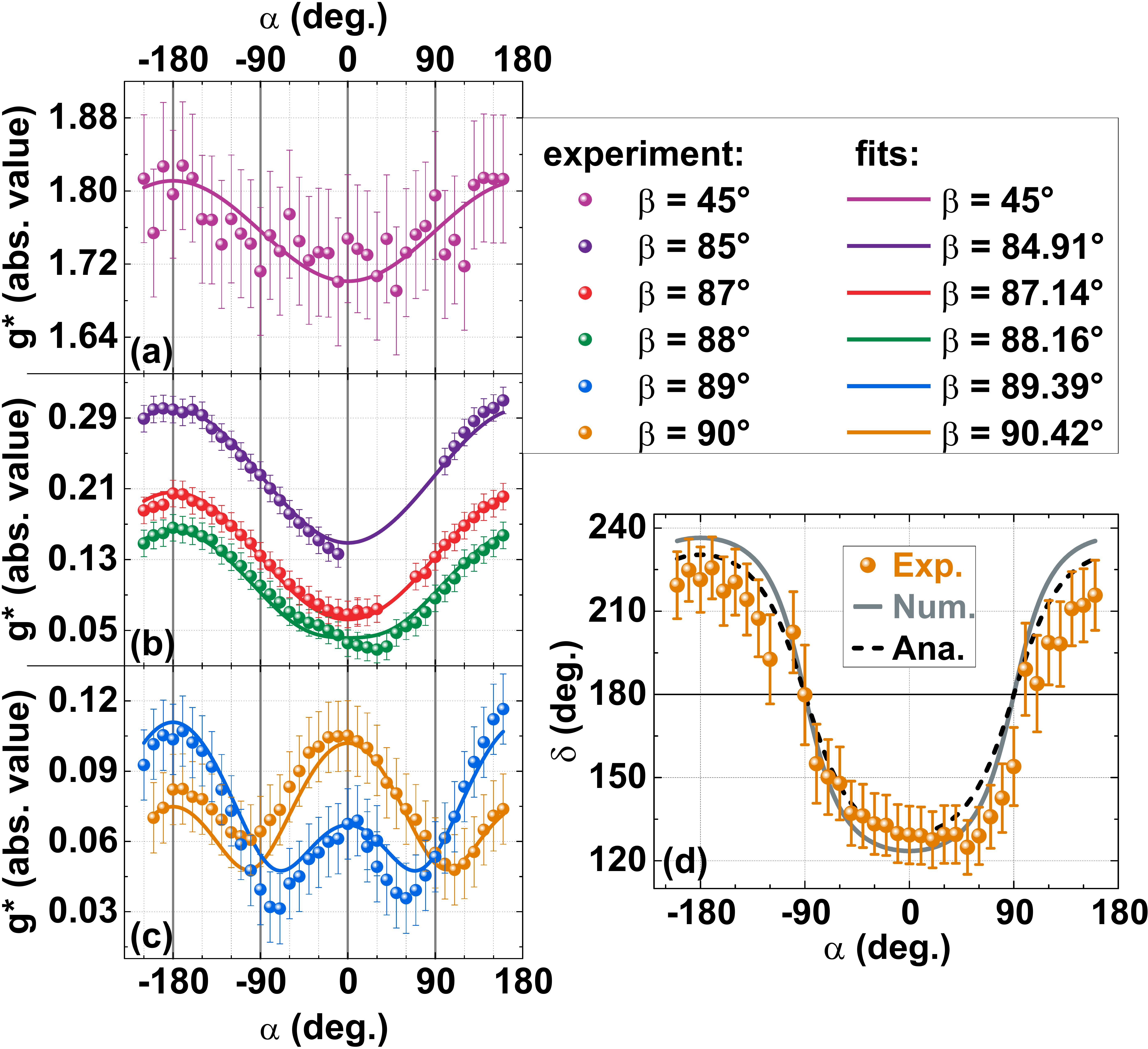}\end{center}
\caption{(a) - (c) $g$ factors gained from TRKR measurements on sample B for different $\beta$ and a complete rotation in $\alpha$. For the fits Eq.\ (\ref{fitformula}) was used. (d) Tilt angle $\delta$ of $\vekc{B}$ out of the QW plane extracted from TRKR measurements compared to the theoretical expectations based on the calculations shown in Secs. \ref{sec:anaresults} and \ref{sec:numcalculations}.}
\label{fig:111_tensor}
\end{figure}

Due to a growth direction slightly tilted away from [111], we expect sample B to show comparable features for the tensor $\vek{g}$ as sample A, though to a lower extent. Therefore, we performed similar TRKR measurements for different values of $\beta$. The extracted effective $g$ factors $g^\ast$ are depicted in Figs.~\ref{fig:111_tensor}(a) to (c). Here, $\alpha = 0^\circ$, i.e., the $x$ axis, and $\alpha = 90^\circ$, i.e., the $y$ axis, correspond to the [$\bar{1}\bar{1}$2] direction and the [$\bar{1}$10] direction, respectively. The effective $g$ factor shows the same behaviour as in sample A, however, the transition from two maxima to one maximum is already visible at around $\beta = 88^\circ$. Additionally, the minima for $\beta = 90^\circ$ are not at $\alpha = \pm90^\circ$.

To be able to fit the data with Eq.\ (\ref{fitformula}), we had to treat $\beta$ as an additional free fit parameter, except for $\beta = 45^\circ$. The extracted values of $\beta$ are slightly different from the expected values, indicating the experimental inaccuracy to perfectly align the sample in the magnetic field. This explains the shifted minima for $\beta = 90^\circ$, too. The extracted parameters are $G_{xx} = 0.00779\pm0.00018$, $G_{zz} = 6.16\pm0.04$, $G_{xz} = -0.192\pm0.003$ and $G_{yy} = 0.00226\pm0.00008$.
Taking again the signs $\det \tilde{\vek{g}}_\mathrm{red} > 0$ and $\tilde{g}_{yy} < 0$ we get
\begin{equation}
  \tilde{\vek{g}} =
  \begin{pmatrix}
  0.0445 & 0 & -0.0762\\
  0 & -0.0475 & 0\\
  -0.0762 & 0 & 2.48
  \end{pmatrix}, \quad \tilde{\delta}(0^\circ)=-59.7^\circ .
\end{equation}
Figure \ref{fig:111_tensor}(d) shows the angle $\delta$ extracted from the in-plane TRKR measurements on sample B compared to the theoretical expectations. Similar to sample A, the sign and phase for the experimental data are adapted to the theoretically calculated sign and phase. Note that the theoretical calculations predict the same $\vek{g}$ tensors for $[mmn]$- and $[\bar{m}\bar{m}\bar{n}]$-oriented growth directions, i.e., Eqs.\ (\ref{eq:g_analytical}) yield the same components $g_{ij}$ for $\theta$ and $\theta'=\theta+180^\circ$. Hence, we calculate $\theta = 57.54^\circ=237.54^\circ-180^\circ$ based on the actual growth direction to discuss and compare the experimental and theoretical $\vek{g}$ tensors. Similar to sample A, a very good agreement between experiment and theory is visible. In $x$ direction the tilting of the effective magnetic field $\vekc{B}$ out of the QW plane is maximum with $\delta(0^\circ)=135^\circ\pm5^\circ$ and $\delta(-180^\circ)=225^\circ\pm5^\circ$ while $\vekc{B}$ is in the QW plane in the $y$ direction with $\delta(\pm90^\circ)\approx180^\circ$. Via $\gamma=195^\circ$ we get
\begin{equation}
  \vek{g} =
  \begin{pmatrix}
  -0.062 & 0 & 0.70\\
  0 & -0.0475 & 0\\
  0.062 & 0 & -2.4
  \end{pmatrix} .
\end{equation}
Similar to sample A we have to consider an error margin of about $\pm7\%$ for the components of $\vek{g}$ except for $g_{yy}$.

The threshold for the transition from two maxima to one maximum yields to $\beta_{0, [111]} = 88.3^\circ$. This leads to a $z'$ axis which is only $1.7^\circ$ tilted away from the $z$ axis. This is in good agreement with the theoretical calculations ($g_{zx}=0$ for [111]-grown QWs), considering a tilt angle of only $2.8^\circ$ of the growth substrate.

\subsection{Tensor $\vek{g}$ in [110]-grown QWs} \label{sec:holegtensor110}

\begin{figure}[t]
\begin{center}\includegraphics[width=8.5cm]{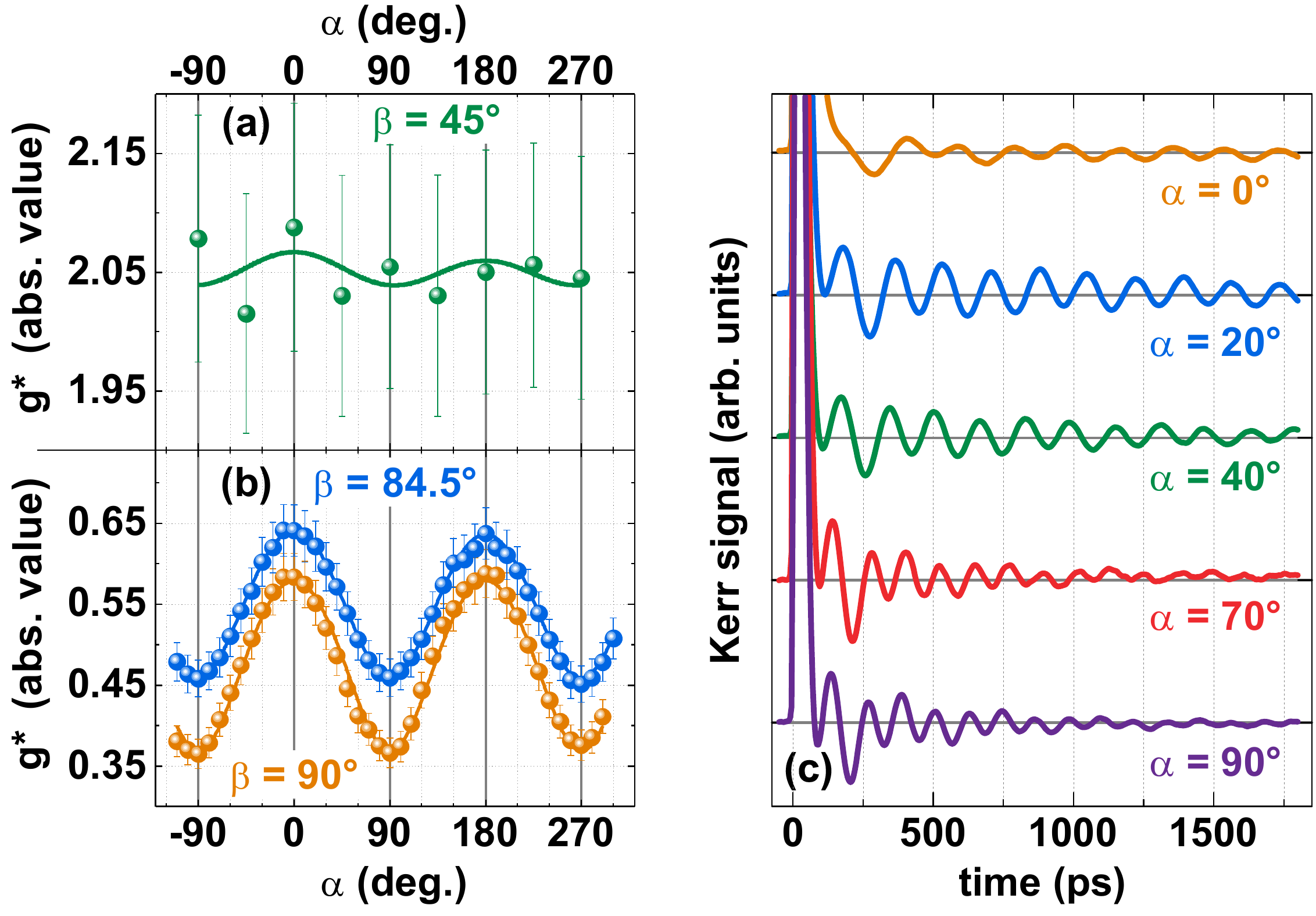}\end{center}
\caption{(a), (b) $g$ factors gained from TRKR measurements on sample C for different $\beta$ and a complete rotation in $\alpha$. For the fit Eq.\ (\ref{fitformula}) was used. (c) Kerr traces measured on sample C for different $\alpha$ and $\beta=90^\circ$.}
\label{fig:110_tensor}
\end{figure}

Similar to the other samples, we analyze the tensor $\vek{g}$ in [110]-grown QWs by performing TRKR measurements for different values of $\beta$, depicted in Figs.~\ref{fig:110_tensor}(a) and (b). Here, $\alpha = 0^\circ$, i.e., $\vek{B}$ along the $x$ axis, and $\alpha = 90^\circ$, i.e., $\vek{B}$ along the $y$ axis, correspond to the [00$\bar{1}$] direction and the [$\bar{1}$10] direction, respectively. The effective $g$ factor for $\beta = 90^\circ$ shows minima at $\alpha = \pm90^\circ$ and $\alpha = 270^\circ$, i.e., $B$ parallel to the $y$ axis, and maxima at $\alpha = 0^\circ$ and $\alpha = 180^\circ$, i.e., $B$ parallel to the $x$ axis. This can be attributed to the in-plane $g$ factor anisotropy, discussed in our previous work \cite{gradl2014}. For $\beta = 84.5^\circ$, the same behaviour can be seen with increased values, indicating a higher out-of-plane $g$ factor. This is supported by the measurements for $\beta = 45^\circ$, which show a very large $g$ factor of about 2. In contrast to the other growth directions, no transition from two maxima to one maximum is visible.

We used Eq.\ (\ref{fitformula}) to fit the data depicted in Figs.~\ref{fig:110_tensor}(a) and (b) and get $G_{xx} = 0.335\pm0.004$, $G_{zz} = 8.18\pm0.02$, $G_{xz} = 0.015\pm0.013$ and $G_{yy} = 0.136\pm0.002$.  Taking the signs $\det \tilde{\vek{g}}_\mathrm{red} > 0$ and $\tilde{g}_{yy} <0$ we get
\begin{equation}
  \tilde{\vek{g}} =
  \begin{pmatrix}
  0.579 & 0 & 0.00432\\
  0 & -0.369 & 0\\
  0.00432 & 0 & 2.86
  \end{pmatrix}, \quad \tilde{\delta}(0^\circ)=0.428^\circ .
\end{equation}
TRKR measurements with in-plane magnetic field [depicted in Fig. \ref{fig:110_tensor}(c)] show no non-oscillatory component for every magnetic field direction. This indicates $g_{zx}=g_{zy}\approx0$ in good agreement with our theoretical calculations. Additionally, a beating of two precession frequencies can be seen (most prominent for $\alpha=90^\circ$), which can be attributed to a combined hole and electron spin precession due to similar effective spin dephasing times for electrons and holes. This makes an accurate quantitative analysis of the non-oscillatory component even more difficult (besides the issues mentioned in Sec.~\ref{sec:holegtensor113}). Therefore, the determination of $\delta=180^\circ\pm20^\circ$ (considering the theoretical predictions) yields a high error margin. Based on $\tilde{\vek{g}}$ and $\gamma=180^\circ$ we get
\begin{equation}
  \vek{g} =
  \begin{pmatrix}
  -0.58 & 0 & -0.026\\
  0 & -0.369 & 0\\
  0 & 0 & -2.9
  \end{pmatrix} .
\end{equation}
Within the error margins the tensor $\vek{g}$ is thus diagonal for the [110] growth direction, in good agreement with our theoretical calculations.

\section{Discussion}
\label{sec:discussion}

\begin{figure}[t]
\begin{center}\includegraphics[width=8.5cm]{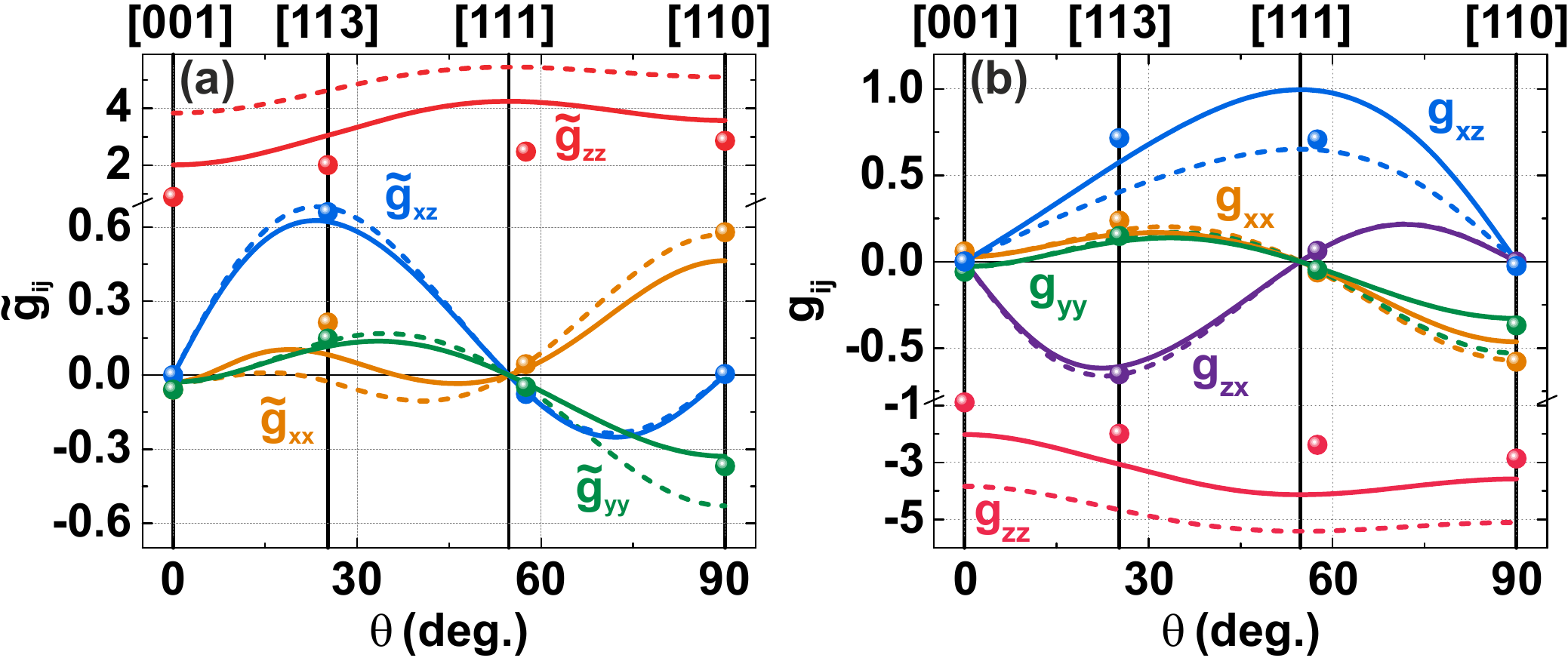}\end{center}
\caption{(a) and (b) Experimental components of $\tilde{\vek{g}}$ and $\vek{g}$ (symbols) compared to numerical (solid lines) and analytical (dashed lines) calculations (for a QW width of $w=12$~nm). The experimental parameters for the [001]-grown QW are taken from Ref.~\onlinecite{korn2010}. \label{fig:discussion}}
\end{figure}

To compare the very accurate experimental data with the theoretical predictions concerning the Zeeman interaction, we derive the components of $\tilde{\vek{g}}$ based on our analytical and numerical calculations (for a QW width of $w=12$~nm) presented in Figs.~\ref{fig:wellwidth}(b)-(d). This is shown in Fig.~\ref{fig:discussion}(a). In addition, experimental parameters for a [001]-grown QW are shown. These components are taken from Ref.~\onlinecite{korn2010}, where, for a 4~nm wide GaAs-QW in Al$_{0.3}$Ga$_{0.7}$As barriers, an in-plane $g$ factor ($g_{xx}$ and $g_{yy}$, respectively) of $|g_\perp|\approx0.05$ and an out-of-plane $g$ factor ($g_{zz}$) of $|g_\parallel|\approx0.89$ was reported (assuming $g_{zx}=g_{xz}=0$).

A very good agreement between the experimentally and theoretically obtained components of $\tilde{\vek{g}}$ can be seen except for $\tilde{g}_{zz}$. The deviation concerning $\tilde{g}_{zz}$ is most likely due to the overestimation of the coupling between the ground subband HH1 to the first excited light hole subband LH2 (mentioned in Sec.~\ref{sec:anaresults}). It is also clearly visible that the numerically calculated components yield yet better agreement with experiment than the analytical model, especially concerning $\tilde{g}_{zz}$.

A comparison of the experimentally gained and theoretically calculated full tensor $\vek{g}$ is depicted in Fig.~\ref{fig:discussion}(b). A very good agreement between experiment and theory can be also seen except for $g_{zz}$. Similar to $\tilde{\vek{g}}$ the numerically calculated components yield best agreement with experiment.

\section{Conclusions}

In conclusion, we have performed low-temperature TRKR measurements of hole spin dynamics to determine the hole $\vek{g}$ tensor in several QWs with different growth directions. We show that the tensor $\vek{g}$ is non-diagonal in QWs grown in the [113] direction as well as in the quasi-[111] direction and is diagonal in QWs grown in the [110] direction.  The peculiar structure of the hole $\vek{g}$ tensor in low-symmetry QWs has drastic consequences for hole spin dynamics: for certain crystallographic orientations, the effective magnetic field driving the spin precession can be almost perpendicular to the externally applied magnetic field.   We analyze our experimental data qualitatively as well as quantitatively and determine the full tensor $\vek{g}$ for the Zeeman interaction. In a theoretical analysis we get explicit analytical expressions as well as accurate numerical results for all components of the tensor $\vek{g}$ for all growth directions $[mmn]$.  A comparison between the experimentally and theoretically gained tensor $\vek{g}$ yields very good agreement.  We show that the tensor $\vek{g}$ is, in general, neither symmetric nor antisymmetric.

For future studies on this topic, different
experimental approaches may be even more suitable to determine the
full tensor $\vek{g}$ without the input from theoretical
calculations. In the current work we were not able to
identify the sign of $g^\ast (\alpha,\beta)$ and therefore we have
adopted the sign of $\det\vek{g}_\mathrm{red}$ and $g_{yy}$ from the
theoretical calculations. However, several approaches have been developped to determine the sign of a scalar $g$ factor experimentally using (time-resolved) luminescence-based techniques \cite{Vekua74, Kalevich1997, Marie00}. More recently, approaches based on TRKR and variations thereof were demonstrated which could be applied to our sample structures. Yang \emph{et al.}\ used non-collinear pump and probe-pulses in a TRKR setup to determine the sign of the $g$ factor \cite{yang2010},  while Kosaka \emph{et al.}\ applied a tomographic Kerr rotation (TKR) method to trace the time evolution of spins in all three dimensions \cite{kosaka2009}. The latter approach would in principle allow the experimental detection of $\vekc{B}$ and therefore $\delta$, which is in our case limited to $\left| \tan \left[ \delta(\alpha) \right] \right|$. We note that, so far, these techniques were only used to observe electron spins with (nearly) isotropic $g$ factors. Therefore, an extension of these experimental techniques to non-diagonal tensors $\vek{g}$ would be beneficial to be able to determine a purely experimental tensor $\vek{g}$.

\section{Acknowledgements}

We acknowledge financial support by the DFG via projects SPP 1285 as well as SFB 689 (project B04) and technical support by Imke Gronwald, Florian Dirnberger and Michael H\"oricke.  RW was supported by the NSF under grant No.\ DMR-1310199. He appreciates stimulating discussions with D.~Culcer and U.~Z\"ulicke.

\bibliographystyle{apsrevtitle}

\begin{thebibliography}{10}
\bibitem{foot:taun}
The parameter $\tau_\mathrm{n}$ can be regarded as the spin relaxation time and
  our measurements yield $\tau_\mathrm{n}\approx\tau_\mathrm{o}$.
\end{thebibliography}

\end{document}